\documentclass{aa}
\usepackage{graphicx}
\usepackage{txfonts}
\begin{document}
%
%   \headnote{Research Note}
   \title{A simple test for periodic signals in red noise}

   \author{S. Vaughan}

   \offprints{S. Vaughan}

   \institute{X-Ray and Observational Astronomy Group, University of
   Leicester,  Leicester, LE1 7RH, U.K.\\
              \email{sav2@star.le.ac.uk}
             }

   \date{Accepted 12/10/2004; Submitted 2004 September 28; in original form  2004 June 10}

   \abstract{
We demonstrate a simple method for testing the  significance of peaks
in the periodogram of  red noise data. 
The procedure was designed to
test for spurious periodicities in X-ray light curves of active
galaxies, but can be used quite generally to test for periodic
components against a background noise spectrum assumed to have a power
law shape.  
The method provides a simple and fast test of the
significance of candidate periodic signals in short, well-sampled time
series such as those obtained from {\it XMM-Newton} observations of Seyfert
galaxies, without the need for Monte Carlo simulations.
A full account is made of the number of trials and the
uncertainties inherent to the model fitting. Ignoring
these subtle effects can lead to substantially overestimated
significances. These difficulties motivate us to 
demand high standards of detection (minimum $>99.9$ per cent confidence) for
periodicities in sources that normally show red noise spectra. 
The method also provides a simple means to
estimate the power spectral index, which may
be an interesting parameter itself, regardless of the presence/absence of
periodicities.

   \keywords{
methods: data analysis -- 
methods: statistical --
X-rays: general --
X-rays: galaxies
               }
   }

   \maketitle

% --------------------------------------------------------------------------
% --------------------------------------------------------------------------
% --------------------------------------------------------------------------
% --------------------------------------------------------------------------

\section{Introduction}
\label{sect:intro}

Many astrophysical sources show erratic, aperiodic brightness
fluctuations with steep power spectra. This type of variability is
known as red noise. By `noise' we mean to say that the intrinsic
variations in the source brightness are random (this has nothing to do with
measurement errors, also called noise). Examples include the X-ray
variability of X-ray binaries (XRBs; e.g. van der Klis \cite{van95})
and Seyfert galaxies (e.g. Lawrence et al. \cite{law87}; Markowitz et
al. \cite{mark}). The power spectrum of these variations, which
describes the dependence of the variability amplitude on temporal
frequency, is often reasonably approximated as a simple power law (over at
least a decade of frequency). This featureless continuum spectrum
does not offer any characteristic frequencies that could be used as
diagnostics.

XRBs often show quasi-periodic oscillations (QPOs) that show-up as
peaks in the power spectrum over the continuum noise spectrum. These
can be thought of as half-way 
between  strictly periodic variations (all power concentrated at one 
frequency) and broad-band noise (power spread over a very broad range
of frequencies). A combination of periodic oscillations with  similar
frequencies, or a single oscillation that is perturbed in frequency,
amplitude or phase can produce a QPO.  QPOs are one of
the most powerful diagnostics of XRB physics (see e.g. van der Klis
\cite{van95}; M$^{\rm c}$Clintock \& Remillard \cite{mr04}).
The detection of periodic or
quasi-periodic variations from a Seyfert galaxy would a be a key
observational discovery, and could lead to a breakthrough in our
understanding if the characteristic (peak) frequency could be
identified with some physically meaningful frequency.
For example, if we assume 
a $1/M_{BH}$ scaling of frequencies we might expect to see
analogues of the high frequency QPOs of XRBs in the range $f_{QPO}
\sim 3 \times 10^{-3} (M_{BH}/  10^6 {\rm M_{\odot}})^{-1}$~Hz 
(Abramowicz et al. \cite{abra}).

However, claims of periodic variations and QPOs in the X-ray emission
of Seyfert galaxies have a chequered history, with no single example
withstanding the test of repeated analyses and observations (see
discussion in Benlloch et al. \cite{ben}).  The confusion arises
partly due to the lack of a standard technique to assess the
significance of a periodicity claim against a background assumption of
random, red noise variability. Indeed, as 
Press (1978) and others have remarked, there is a tendency for the eye to
identify spurious, low frequency periods in random time series.  
Tests for the presence of periodic variations
against a background of white (flat spectrum) noise are well
established, from Schuster (\cite{sch98}) and Fisher (\cite{fis}),
these are reviewed in section 6.1.4 of Priestley
(\cite{pri81}), and discussed in an astrophysical context by 
Leahy et al. (\cite{lea}) and van der Klis (\cite{van89}).
But without modification these methods cannot be used to 
test against red noise variations. 
Timmer \& K\"onig (\cite{tim}) and Benlloch et al. (\cite{ben}) 
have proposed Monte Carlo testing methods applicable to red noise but
the relatively high computational demands of these methods may be
enough to deter some potential users. Israel \& Stella
(\cite{is96}) proposed a method that does not require Monte Carlo 
simulations but is not optimised for short observations of power law
continuum spectra.

This paper puts forward a simple test that can be used to test the significance 
of candidate periodicities superposed on a red noise spectrum which has
an approximately power law shape. The price of simplicity, in this
case, is that the test is only strictly valid when the
underlying continuum spectrum is a power law. The basic steps of the method are:
($i$) measure the periodogram, ($ii$) estimate the red noise continuum
spectrum and ($iii$) estimate the significance of any peaks above the
continuum. The stages of the method are explained in detail in the
following sections. Section~\ref{sect:per} gives a
brief introduction to the statistical properties of the periodogram.
Section~\ref{sect:fitting} discusses a simple method for estimating
the parameters of a power law-like spectrum and section~\ref{sect:stats}
discusses how to estimate the significance of a peak above the
continuum. 
Section~\ref{sect:sims} then demonstrates the veracity of the
method using Monte Carlo simulations
and section~\ref{sect:caveats} reviews some important caveats that must
be considered when using this (and other) period-searching methods.
Finally, section~\ref{sect:disco} gives a brief review of the method
in the context of observations of active galaxies. The appendix
discusses a more generally applicable method of periodogram fitting
(that makes no assumption on the form of the underlying spectrum).

% --------------------------------------------------------------------------
% --------------------------------------------------------------------------
% --------------------------------------------------------------------------
% --------------------------------------------------------------------------

\section{The periodogram}
\label{sect:per}

Given an evenly sampled time series $x_k$ of $K$ points sampled at
intervals $\Delta T$ we can measure its
periodogram (Schuster \cite{sch98}), which is simply the
modulus-squared of the discrete Fourier transform, $X(f_j)$, at each
of the $n=K/2$ Fourier frequencies:
\begin{equation}
\label{eqn:ft2}
I(f_j) = \frac{2\Delta T}{\langle x \rangle^2 N} |X_j|^2 
\end{equation}
The normalisation is chosen such that the units of the periodogram 
are  (rms/mean)$^{2}$ Hz$^{-1}$ (where rms/mean is dimensionless) and
summing the periodogram over positive frequencies gives the sample
variance in fractional units. 
The periodogram is evaluated at the Fourier frequencies 
$f_j = j / K \Delta T$ with $j=1,2,\ldots,K/2$.
The original purpose of the
periodogram was as a tool for  identifying `hidden periodicities' in
time series.  However, the periodogram of a noise process, if measured
from a single time series, shows a  great deal of scatter around the
underlying power spectrum.   Specifically, the periodogram at a given
frequency, $I(f_j)$, is scattered around the true power spectrum,
$\mathcal{P}(f_j)$, following a $\chi^{2}$ distribution with two
degrees of freedom:
\begin{equation}
\label{eqn:pds_scatter}
I(f_j) = \mathcal{P}(f_j) \chi_{2}^{2}/2,
\end{equation}
where $\chi_{2}^{2}$ is a random variable distributed as $\chi^{2}$
with two degrees of freedom, i.e. an exponential probability
distribution with a mean and variance of two and four, respectively:
\begin{equation}
p_{\chi^{2}}(x) = {\rm e}^{-x/2}/2  
\label{eqn:chi2}
\end{equation}
The periodogram is distributed
in this way  because the real and imaginary parts of the DFT are
normally distributed for a stochastic process\footnote{ The DFT at the
Nyquist frequency is always real when $N$ is even so the periodogram
at this frequency  is distributed as $\chi_{1}^{2}$, i.e. with one
degree of freedom.}, and the sum of two squared normally distributed
variables is a $\chi_{2}^{2}$-distributed variable (Jenkins \& Watts
\cite{jen}; Priestley \cite{pri81}; Chatfield \cite{cha}; Bloomfield
\cite{blo}). See also Scargle (\cite{sca82}), Leahy et
al. (\cite{lea}), Wall \& Jenkins (\cite{wj}) and Groth (\cite{gro})
for further discussion of this point. 

If the spectrum is flat (`white noise') and its power level known {\it
a priori} then we can simply make use of the known probability
distribution (equations~\ref{eqn:pds_scatter} and \ref{eqn:chi2})
to  estimate the likelihood that a given periodogram
ordinate exceeds some threshold. If the power level is not known there
is an added uncertainty, but nevertheless an  exact test does exist ({\it
Fisher's g statistic}: Fisher \cite{fis}; section 6.1.4 of Priestley
\cite{pri81}) to estimate the likelihood that the highest peak in the
periodogram was caused by a random fluctuation in the noise spectrum
(see also Koen \cite{koe}).

\begin{figure}
\centering
\includegraphics[width=6.40 cm, angle=270]{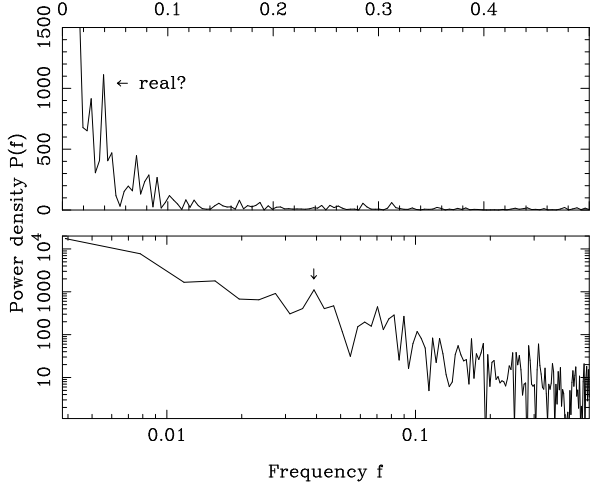}
\caption{
Periodogram of a short ($K=256$) time series containing
red noise.
The upper panel shows the periodogram using linear axes,
the lower panel shows the same data on logarithmic axes.
The periodogram shows a red noise spectrum
rising at lower frequencies. But the periodogram also 
shows a peak at $f=4\times 10^{-2}$. Is this due to a
real harmonic (periodic) variation or an artifact of
the fluctuating noise spectrum?
\label{fig:test_qpo_psd}}
\end{figure}

For the more general case of non-white noise there is no such exact
test.  When examining the periodogram of red noise data such as from
Seyfert galaxies, we need to be careful not to identify spurious
peaks.  Even in the `null' case (i.e. no periodic component)  peaks
may occur in the periodogram due to sampling fluctuations.  In
particular the eye may be drawn to low frequency peaks because, in red
noise data, there is much more power and more scatter in the
periodogram at low frequencies.  Given a large amount of data we can
average the periodogram in  one of the standard ways (see e.g. van der
Klis \cite{van89}; Papadakis \& Lawrence \cite{pap}), fit the
continuum using a standard $\chi^2$-minimisation tool (e.g. Bevington
\& Robinson \cite{bev}; Press et al. \cite{press96}) and test of the
presence of addition features.  This is the standard procedure for
analysing XRB data.  If we have a very limited amount of data, such
that we cannot afford to average the periodogram, we are faced with a
more difficult  situation. Figure~\ref{fig:test_qpo_psd} gives an
example like this. The periodogram of a short time series, containing
red noise (generated using the method of Timmer \& K\"onig
\cite{tim}), shows a large peak at $f=4\times 10^{-2}$.  Could this be
due to a real periodic variation present in the data or is it just a
fluctuation in the red noise spectrum?    

% --------------------------------------------------------------------------
% --------------------------------------------------------------------------
% --------------------------------------------------------------------------
% --------------------------------------------------------------------------

\section{Fitting the periodogram}
\label{sect:fitting}

\subsection{Least squares (LS) fit to log-periodogram}

If the underlying power spectrum is suspected to be a power law then
then parameters of interest are its slope, $\alpha$, and normalisation
$N$.  One of the simplest methods to estimate these parameters from the
raw (unbinned) periodogram is to fit it with a model of the form
$\mathcal{P}(f)=Nf^{-\alpha}$ using the method of least squares (LS).  The
problem with this is that the periodogram is distributed around the
true underlying spectrum in a non-Gaussian fashion and, more
seriously,  the distribution depends on the spectrum itself
(equation~\ref{eqn:pds_scatter}).

To simplify the problem we can fit the logarithm of
the periodogram, as discussed in some detail by Geweke \& Porter-Hudak
(\cite{gph}; see also Papadakis \& Lawrence \cite{pap}). The scatter
in the periodogram scales with the spectrum  itself; the scatter is
multiplicative in linear-space.   This means the scatter is additive
in log-space:
\begin{equation}
\log[I(f_j)] = \log[\mathcal{P}(f_j)] + \log[ \chi_{2}^{2}/2]
\end{equation}
and therefore identical at each frequency (the data are {\it
  homoskedastic}). 
Working with the logarithm of the periodogram has two further
advantages. The first is that  if the power 
spectrum is a power law, the natural way to plot the data is in log
space. The power law becomes  a  linear function: $\log [
\mathcal{P}(f) ] = \log [ N ] - \alpha \log [ f ] $. The second
advantage is that the distribution of periodogram ordinates becomes
less skewed.  This reduces the effect of `outliers' on the fitting.

We must be careful fitting the logarithm.  The expectation value of
the logarithm of the periodogram is \emph{not} the expectation value
of the logarithm of the spectrum. However, the bias is a constant (due
to the shape of the $\chi_{2}^{2}$-distribution in log-space) that can
be removed trivially:
\begin{equation}
\langle \log[I(f_j)] \rangle = \langle \log[\mathcal{P}(f_j)] \rangle
+  \langle \log[ \chi_{2}^{2}/2] \rangle
\end{equation}
Using equation 26.4.36 of Abramowitz \& Stegun (\cite{abr}) we have
$\langle \log[ \chi_{2}^{2}/2] \rangle = -  0.57721466\ldots/\ln
[10]$. (The number in the numerator is Euler's constant.) Therefore
\begin{equation}
\langle \log[\mathcal{P}(f_j)] \rangle =  \langle \log[I(f_j)]\rangle
+0.25068 \ldots
\end{equation}
The logarithm of the periodogram ordinate, with the bias removed
(i.e. the $0.25068$ added\footnote{ This is a more precise
approximation to $\langle \log[ \chi_{2}^{2}/2] \rangle$ than used by
Papadakis \& Lawrence (\cite{pap}).  }), is thus an unbiased estimator
of the logarithm of the spectrum, and is distributed independently and
identically (about the underlying spectrum) at each frequency.   (The
raw periodogram points are not identically distributed in linear space
since the scatter depends on the spectrum, which is a function of
frequency.)  Thus we can use a LS fitting procedure 
to get a reasonable estimate of the power spectral slope $\alpha$ and
normalisation $N$ by
fitting a linear function $y=mx+c$ to the plot of $\log[I(f_j)]$
versus $\log[f_j]$. The slope of the linear fit gives $\hat{\alpha}=
-m$ and the $y$-intercept gives $\log(\hat{N}) = c + 0.25068$.
These give our estimate for the continuum: $\hat{\mathcal{P}}_j
= \hat{N} f_j^{-\hat{\alpha}}$, or equivalently $\log[\hat{\mathcal{P}}_j]
= \log[ \hat{N} ] - \hat{\alpha} \log[ f_j ]$.

\begin{figure}
\begin{center}
   \includegraphics[width=6.4 cm, angle=270]{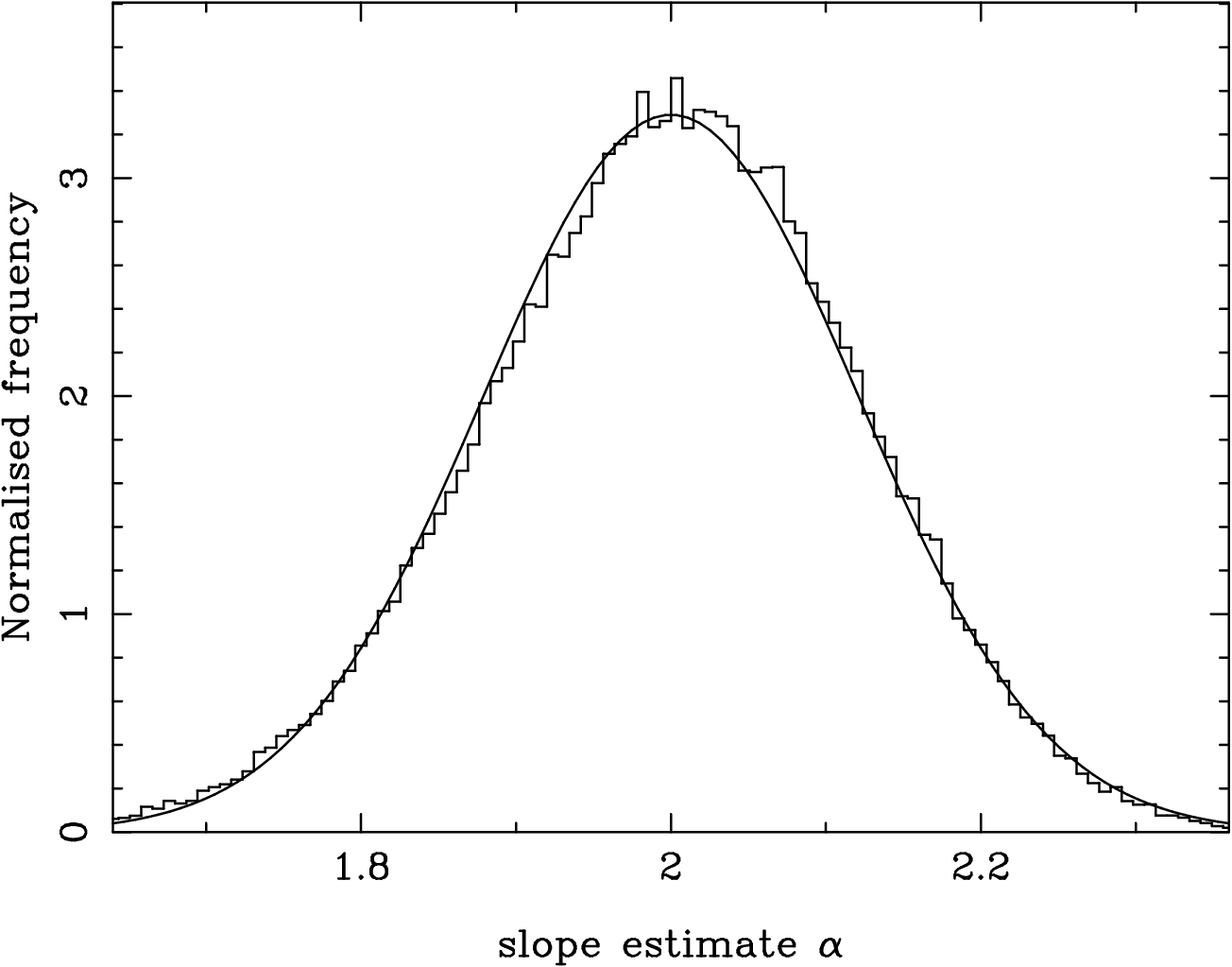}

\vspace{0.5 cm}
   \includegraphics[width=6.4 cm, angle=270]{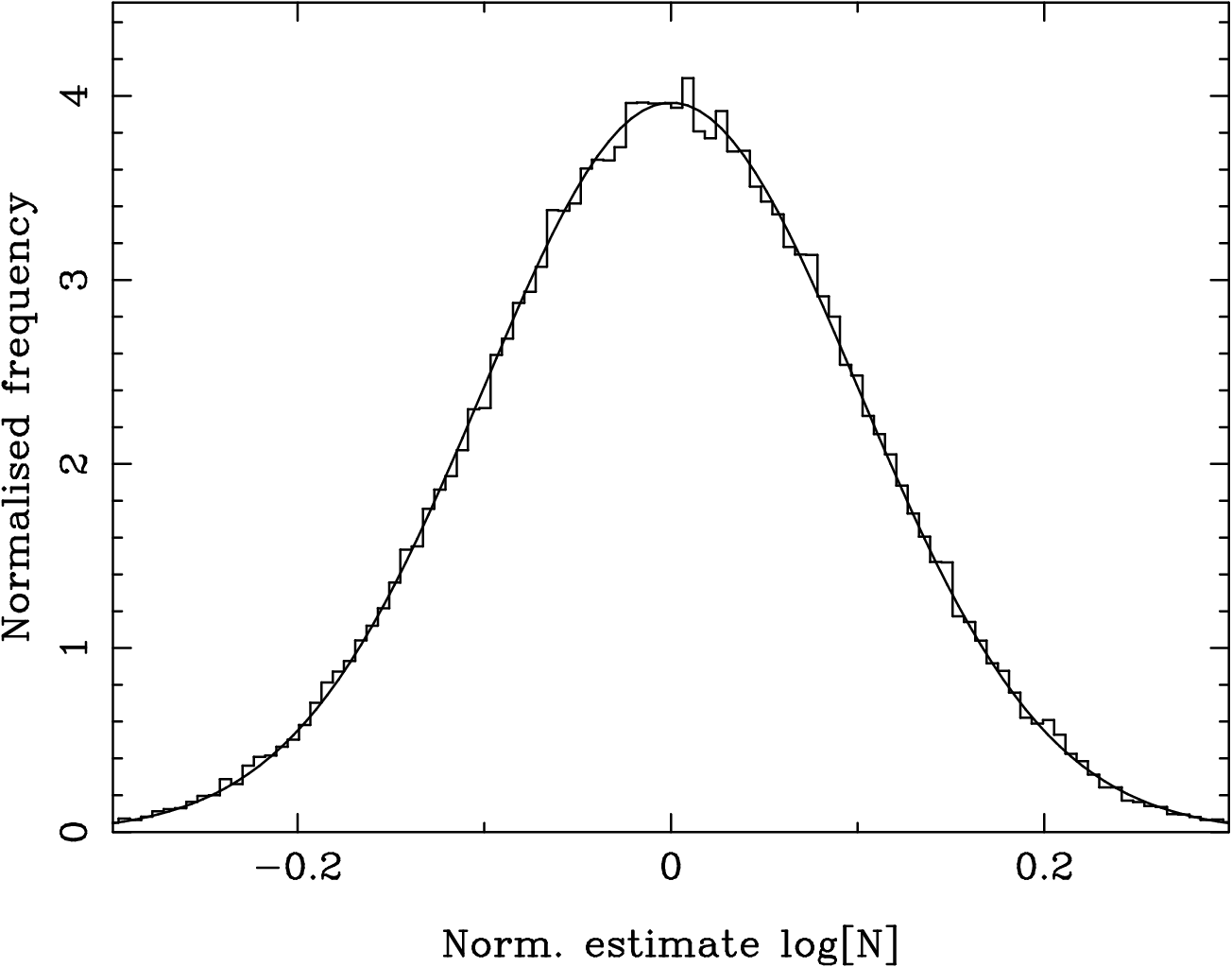}
\end{center}
\caption{
Distribution of the slope and normalisation estimators derived from
$10^5$ Monte Carlo simulations of $K=256$ point time series
(histogram). The `true' spectral parameters were $\alpha=2$ and $\log[N]=0$. 
The predictions of Gaussian uncertainties, with widths given by equations
~\ref{eqn:slope_err} and \ref{eqn:norm_err}, are shown with the
smooth curves.
\label{fig:LS_dist}
}
\end{figure}

It is important to note that the datum at the Nyquist frequency 
($j=n$) should be ignored in the LS fitting. This is because, 
as mentioned previously, the distribution of the periodogram ordinate
at this frequency is not identical to that at other frequencies (it follows
a $\chi_1^2$ distribution). This minor detail means that the
LS fit should be performed on the $n^{\prime} = n-1$ lowest frequencies 
that are identically distributed (in log-space).

A drawback of fitting the periodogram, rather than
the binned or averaged periodogram, is that it does not
provide a in-built goodness-of-fit test. By binning
the periodogram (van der Klis \cite{van89}; Papadakis \& Lawrence
\cite{pap}) we can obtain Gaussian errors on each ordinate to be
used in a $\chi^2$-test. We do not have Gaussian error bars for
the unbinned log-periodogram. But, since we know the expected distribution
of the periodogram ordinates about the true spectrum we can
compare this to the distribution of residuals from the fitted
data using a Kolmogorov-Smirnov test (Press et al. \cite{press96}). 
Specifically, we can compare the data/model ratio (in linear space)
given by $\hat{ \gamma}_j = 2 I_j / \hat{\mathcal{P}}_j$
with the theoretical $\chi_2^2$ distribution, if the model is
reasonable $\hat{\gamma}_j$ should be consistent with the $\chi_2^2$
distribution. Furthermore, the KS test is most sensitive around the
median value, and less sensitive at the tails of the distribution,
which means that even in the presence of a real periodic signals
(i.e. a few outlying powers)
the test should give a good idea of the overall quality of the
continuum fit.

% --------------------------------------------------------------------------

\subsection{Uncertainties on the parameters}
\label{sect:error}

The uncertainties on the slope and normalisation estimates from 
the LS method can be derived using the standard theory of linear regression 
(e.g. Bevington \& Robinson \cite{bev}; Press et al. \cite{press96}).
The error on the slope (index) and intercept (log normalisation) are:
\begin{equation}
err^2 [ \hat{\alpha} ] = \frac{n^{\prime} \sigma^2}{ \Delta }
\label{eqn:slope_err}
\end{equation}
and 
\begin{equation}
err^2 [ \log(\hat{N}) ] = \frac{ \sigma^2 \sum_{j=1}^{n^{\prime}} a_j^2  }{ \Delta }
\label{eqn:norm_err}
\end{equation}
where $a_j= \log[f_j]$ and
\begin{equation}
\Delta = n^{\prime} \sum_{j=1}^{n^{\prime}} a_j^2 - \left( \sum_{j=1}^{n^{\prime}} a_j \right)^2  
\end{equation}
and also $\sigma^2=\pi^2/6 (\ln [10])^2$ is the variance of the
log-periodogram ordinates about the true spectrum (Geweke \&
Porter-Hudak \cite{gph}). The covariance of the two parameter estimates
is given by:
\begin{equation}
cov(\hat{\alpha},\log[\hat{N}]) = \frac{ \sigma^2 \sum_{j=1}^{n^{\prime}} a_j }{ \Delta }
\label{eqn:covar}
\end{equation}
Here $n^{\prime}$ is the number of frequencies used in the fitting.
Normally $n^{\prime}=n-1$ since only the Nyquist frequency is
ignored (because the periodogram at the Nyquist
frequency does not follow the same distribution as at the other
frequencies). 
%Since the frequencies $f_j$ are just the linearly spaced Fourier
%frequencies ($f_j = j / K \Delta T$)) these expressions can be further
%approximated. For $K > 100$ the errors can be roughly approximated as $ err [
%\hat{\alpha} ] \sim 2 / \sqrt{K}$, $err[\log(\hat{N})] \sim 2 /
%\sqrt{K}$ and $cov[\alpha,\log(\hat{N})] \sim - 2 / K$. 

\begin{figure}
\begin{center}
   \includegraphics[width=6.4 cm, angle=270]{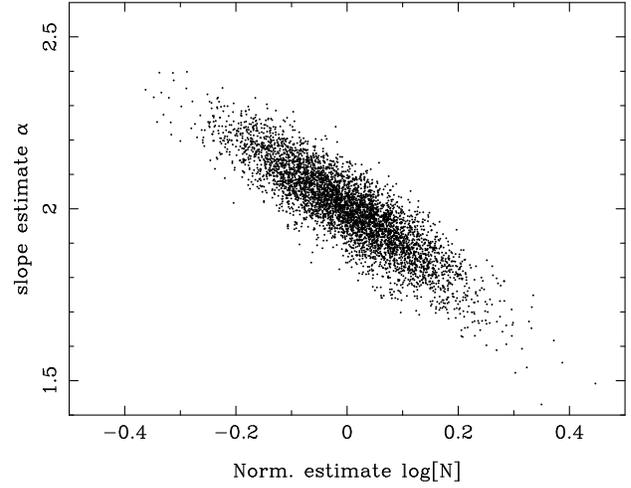}
\end{center}
\caption{
Demonstration of the covariance in the 
estimates of slope and normalisation from 
from LS fitting
to the logarithm of the periodogram. 
The plot shows the results of fitting $5,000$ 
Monte Carlo simulations of an $\alpha=2$, $N=1$ spectrum.
\label{fig:LS_cov}
}
\end{figure}

The accuracy of these equations was tested using a Monte Carlo simulations.  An
ensemble of random time series, each of length $K$, was generated
(using the method of  Timmer \& K\"onig \cite{tim}). For each series
the power spectral slope and normalisation were estimated using the
LS method discussed above. Figure~\ref{fig:LS_dist} shows the distribution of
estimates for $10^5$ realisations of time series 
generated by a process with an $\alpha=2, N=1$ spectrum. With only
$n=127$ periodogram points (i.e. $K=256$) the distribution of the estimates is
reasonably close to Gaussian.    (The
distribution of $\hat{N}$ is log-normal because the estimated
quantity $\log[\hat{N}]$ is normally distributed in the LS fitting.)
These two parameters are covariant in the fit; a low estimate of the
slope tends to be correlated with a high estimate of the normalisation.
Figure~\ref{fig:LS_cov} illustrates the covariance between the
two estimated parameters.
The shape of these distributions is independent of the
spectral slope, this was confirmed using
Monte Carlo simulations of spectra
with slopes in the range $\alpha=0-3$.

\begin{figure}
\begin{center}
   \includegraphics[width=8.0 cm, angle=0]{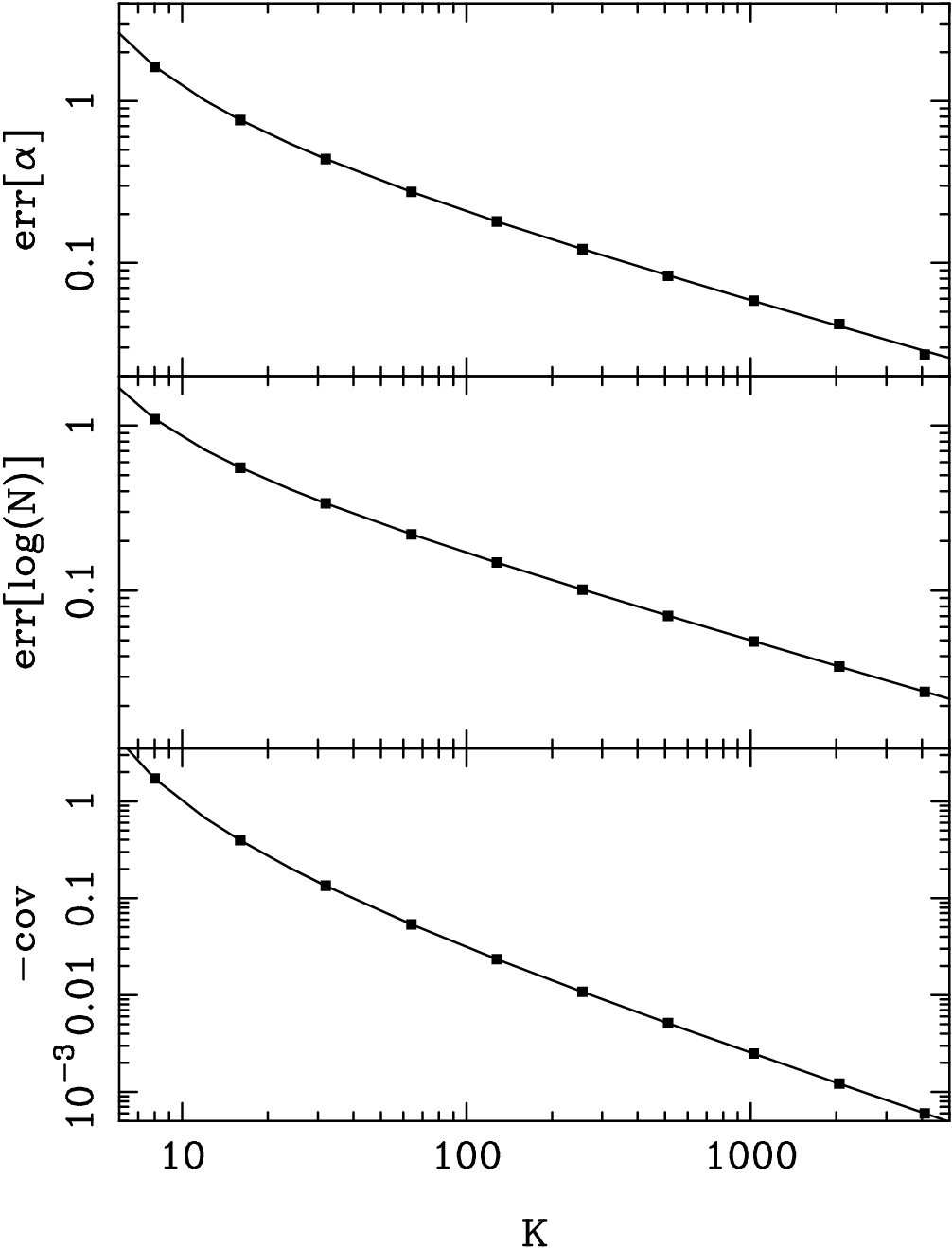}
\end{center}
\caption{
Demonstration of the uncertainties from LS fitting
to the logarithm of the periodogram. 
For each value of $K$, $10^5$ time series 
were simulated (with an $\alpha=2$, $N=1$ spectrum).
The periodogram of each time series was fitted (in log space)
with a linear model. From the $10^5$ estimates of the two 
parameters (slope and normalisation) their rms and covariance were measured 
(solid squares). The solid lines mark the predictions of
equations~\ref{eqn:slope_err}, \ref{eqn:norm_err} and \ref{eqn:covar}.
%The dotted lines denote the approximate error expressions
%$err[\hat{\alpha}] \sim 2 / \sqrt{K}$, $err[\log(\hat{N})] \sim  2 /
%\sqrt{K}$ and $cov[\alpha,\log(\hat{N})] \sim - 2 / K$. 
\label{fig:LS_err}
}
\end{figure}

The uncertainties on $\hat{\alpha}$ and $\log[\hat{N}]$, and their
covariance, were estimated for different length series 
by the same Monte Carlo method as discussed above.
These Monte Carlo uncertainties compare well with the theoretically
expected uncertainties for the LS fitting method as discussed above
(Fig.~\ref{fig:LS_err}). 

% --------------------------------------------------------------------------

\subsection{Uncertainty on the model}
\label{sect:error2}

The expected uncertainties and covariance of the two model parameters
can be combined to give an estimate of the uncertainty
of the logarithm of the model, $\log [ \hat{\mathcal{P}}_j ]$, at a
frequency $f_j$, using the standard error propagation formula.
\begin{eqnarray}
err^2 [\log \{ \hat{\mathcal{P}}(f_j) \} ] & = &
err^2 [ \hat{\alpha} ] \times (\log[f_j])^2 +
err^2 [\log(\hat{N})] - \nonumber \\
 & & 2 cov[\alpha,\log(\hat{N})] \times (\log[f_j])
\label{eqn:model_err}
\end{eqnarray}
The first term in the sum accounts for 
the uncertainty on the slope (equation~\ref{eqn:slope_err}), 
the second accounts for the uncertainty in the normalisation (equation~\ref{eqn:norm_err})
and the third accounts for their covariance (equation~\ref{eqn:covar}). 
The uncertainty is frequency dependent. Even in log-space the
model is more uncertain at lower frequencies,
this simply reflects the fact that there are many more points
at high $\log[f]$ than at low $\log[f]$. 
But also note that the uncertainty on the logarithm of the model
is independent of the model itself (slope and normalisation).

\begin{figure}
\begin{center}
   \includegraphics[width=6.4 cm, angle=270]{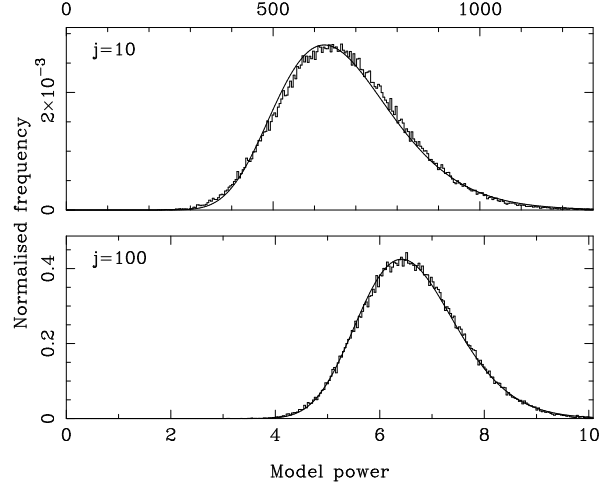}
\end{center}
\caption{
Monte Carlo demonstration of the distribution of power
in the model (power law) spectrum.
Using an $\alpha=2$, $N=1$ spectrum, $10^5$ random time series
of length $K=256$ were generated. For each one the 
periodogram was fitted as discussed in the text. 
The histograms show the distribution of the power in the resulting
$10^5$ models at the $j=10$ and $j=100$ Fourier
frequencies ($f_j=j/K\Delta T$). The solid line shows the predicted
log-normal distribution (equation~\ref{eqn:lognorm}).
\label{fig:model_err}
}
\end{figure}

The distribution of the power
in the model (in log-space), $\log [\mathcal{P}_j]$, 
is expected to be Gaussian with a width
determined by the formula above. In linear-space the
uncertainty on the model power, $\mathcal{P}_j$, is
log-normally distributed. The probability density function for
the model power is therefore
\begin{equation}
p_{\hat{\mathcal{P}}_j}(y) = \frac{1}{S_j y \sqrt{2 \pi}} 
\exp \left\{ - \frac{(\ln[y] - M_j)^2}{2S_j^2}  \right\}
\label{eqn:lognorm}
\end{equation}
where $M_j=\ln[\mathcal{P}_j]$ is the expected value of
the power (in log-space) and $S_j$ is the rms width of the
distribution of powers (also in log-space) as given by
equation~\ref{eqn:model_err}. Both of these are
frequency dependent. Note that the log-normal
distribution is conventionally defined in terms of the natural
logarithm, whereas 
previously the results were given in terms of base $10$
logarithm. The uncertainty on the model log-powers 
from equation~\ref{eqn:model_err} needs to be corrected:
\begin{equation}
S_j = err[ \log \{ \hat{\mathcal{P}}(f_j) \} ] \times \ln[10]
\label{eqn:model_err2}
\end{equation}
Figure~\ref{fig:model_err} shows the distribution
of model powers (at two different frequencies) for
$10^5$ Monte Carlo simulations of $K=256$ time series.
Clearly the predicted log-normal distribution (with parameters $M_j$ and
$S_j$ as given above) gives a good description of the real uncertainty
in the model.

% --------------------------------------------------------------------------

\subsection{Bias in the parameters}

\begin{figure}
\begin{center}
   \includegraphics[width=8 cm, angle=0]{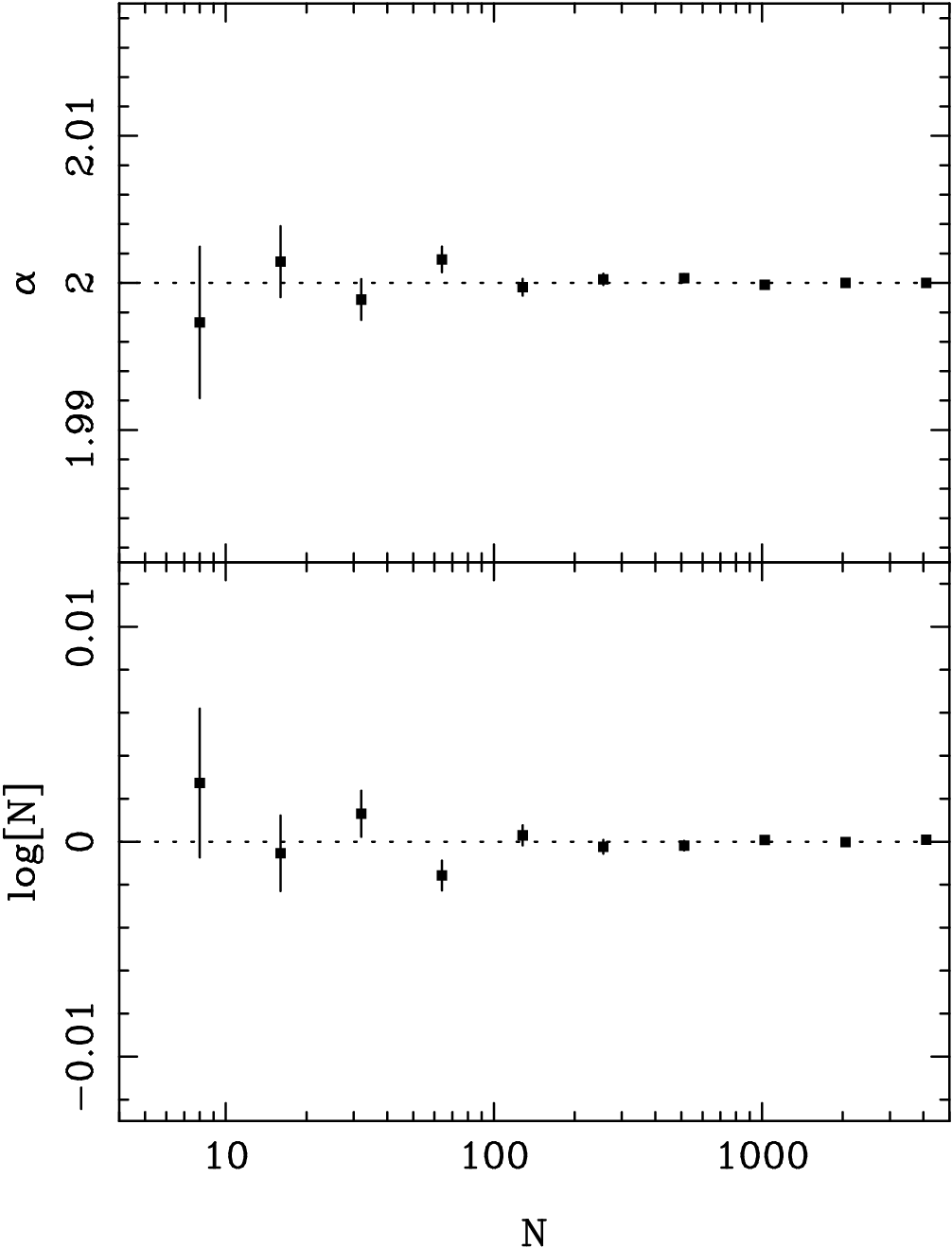}
\end{center}
\caption{
Demonstration of bias in LS fitting
of the logarithm of the periodogram. 
For each value of $K$, $10^5$ time series 
were simulated (with an $\alpha=2$, $N=1$ spectrum) and
the periodogram of each time series was fitted (in log space; see text)
with a linear model. The parameters estimates
were averaged over the $10^5$ realisations. 
The estimators $\hat{\alpha}$ and $\log[\hat{N}]$ are 
clearly unbiased.
\label{fig:LS_bias}
}
\end{figure}

Although in general the LS method does not yield
the maximum likelihood solution for non-Gaussian data,
for the specific case of a power law spectrum the
parameters obtained from the log-periodogram regression,
namely $\hat{\alpha}$ and $\log[\hat{N}]$, are unbiased.
Figure~\ref{fig:LS_bias} demonstrates this using 
Monte Carlo simulations. However, it should be noted that because the
parameter $\log[\hat{N}]$ is normally distributed the parameter
$\hat{N}$ will be log-normally distributed. Thus the mean value of
$\hat{N}$ is not a good estimator of the true value (it will be
biased upwards due to the long tail of the log-normal distribution). 

% --------------------------------------------------------------------------

\subsection{Summary of LS fitting}

The following summarises the LS fitting method:
\begin{itemize}

\item
Measure the periodogram of the time series.

\item
Ignore the Nyquist frequency point.

\item
Take the logarithm.

\item
Fit a straight line using the standard LS method.

\item
Test the goodness-of-fit using the KS test.

\item
Estimate uncertainties on the parameters and the model.

\end{itemize}

The uncertainties and covariance of the parameters can be
computed using equations~\ref{eqn:slope_err}, \ref{eqn:norm_err} and
\ref{eqn:covar}. 
%(or the approximations discussed in section~\ref{sect:error})
These can then be used to calculate
the uncertainty on the logarithm of model power $S_j$
(equation~\ref{eqn:model_err2}). 

% --------------------------------------------------------------------------
% --------------------------------------------------------------------------
% --------------------------------------------------------------------------
% --------------------------------------------------------------------------

\section{Confidence limits}
\label{sect:stats}

\subsection{The ideal case}
\label{sect:ideal_stats}

If we know the exact form of the spectrum we can divide this out of the
periodogram. From equation~\ref{eqn:pds_scatter} we can see that the
ratio 
$\gamma_j \equiv 2 I(f_j) /  \mathcal{P}(f_j)$  will be distributed
like $\chi^2_2$. We can
use our estimates $\hat{\alpha}$ and $\hat{N}$ to form the null
hypothesis: the data were generated by a process with a spectrum
$\hat{\mathcal{P}}(f_j)=\hat{N}f_j^{-\hat{\alpha}}$ and no periodic component.
We can estimate the probability that a large peak will occur
in the periodogram, assuming the model spectrum, by comparing
$\hat{\gamma}_j$ to the $\chi^2_2$ PDF (Priestley 1981; Scargle 1982).

We can define a $(1-\epsilon)100$ per cent confidence limit on $\hat{\gamma}$,
call this $\gamma_{\epsilon}$, as the level for which, at a given frequency, the
probability of obtaining a higher value by chance is $\Pr \{ \hat{\gamma} >
\gamma_{\epsilon} \} = \epsilon$ on the assumption that the null hypothesis is
true. The chosen value of $\epsilon$ represents the `false alarm
probability'.
The integral of the $\chi_2^2$ probability distribution  gives
the probability of a single sample exceeding a value of $\gamma_{\epsilon}$ by
chance:
\begin{equation}
\Pr \{ \chi^2 > \gamma_{\epsilon} \} = \frac{1}{2} \int_{\gamma_{\epsilon}}^{\infty} {\rm e}^{ -x/2
} dx  = {\rm e}^{-\gamma_{\epsilon}/2} = \epsilon
\end{equation}
For a given probability $\epsilon$, we can rewrite this: 
\begin{equation}
\gamma_{\epsilon} = - 2 \ln[\epsilon].
\label{eqn:1-error}
\end{equation}
For example, using $\epsilon=0.05$ (i.e. a $95$ per cent significance
test) we find $\gamma_{0.05} = 5.99$.  This means
that if the null hypothesis is true the probability of the
ratio $\hat{ \gamma_j} $ being higher than $5.99$ is only $0.05$.  
We can therefore define our $95$ (and $99$) per cent confidence limits
on the log-periodogram as the model
$\hat{\mathcal{P}}(f)=\hat{N}f^{-\hat{\alpha}}$ 
multiplied by the appropriate $\gamma_{\epsilon}/2$. (In log-space we
simply add the appropriate $\log[ \gamma_{\epsilon}/2] $ to the model.)

However, these confidence bounds correspond to a \emph{single trial}
test, they give the probability that a periodogram point at one particular
frequency will exceed 
$\gamma_{\epsilon}\mathcal{P}_j/2$.  Usually there are $n^{\prime} = n - 1$
independent trials since only the Nyquist frequency
is ignored (leaving $n^{\prime}$ independently distributed
periodogram points to be examined). 
We must account for the number of independent trials:
\begin{equation}
\gamma_{\epsilon} = - 2 \ln [ 1 - ( 1 - \epsilon_{n^{\prime}} )^{1/n^{\prime}} ]
\label{eqn:global}
\end{equation}
This gives the value of $\hat{\gamma}$ that has a
probability of being exceeded of $\epsilon_{n^{\prime}}$ in
$n^{\prime}$ independent 
trials. In the limit of large $n^{\prime}$ and small $\epsilon$ this can
be approximated as $\gamma_{\epsilon} = - 2 \ln
[\epsilon_{n^{\prime}}/n^{\prime}]$. Of course, if we knew which
frequencies to test \emph{a priori} then we could perform $<n^{\prime}$
independent tests and the significance levels could be adjusted
accordingly.

% --------------------------------------------------------------------------

\subsection{Accounting for model uncertainty}
\label{sect:model_err}

The case outlined above is valid only when we know the true power
spectrum exactly ($\hat{\mathcal{P}}_j=\mathcal{P}_j$). 
In reality all we have
is an  estimated model $\hat{\mathcal{P}}_j$ (which will differ from
the true spectrum) and its uncertainty.  This extra
uncertainty alters the probability distribution.  The ratio $\hat{\gamma}_j
= 2 I_j / \hat{\mathcal{P}}_j$ is really the ratio of two random
variables; the PDF of this would allow us to calculate the probability
of observing a given value of $\hat{\gamma}_j$ taking full account of
the uncertainty in the model fitting. As discussed above $2I_j$ will 
follow a rescaled $\chi_2^2$ distribution about the true spectrum. 
\begin{equation}
p_{2I_j}(x) = \frac{1}{2 \mathcal{P}_j} {\rm
  e}^{-x/2 \mathcal{P}_j} 
\end{equation}

In the case of the LS fitting discussed in
section~\ref{sect:fitting} the model $\hat{\mathcal{P}}_j$ has a
log-normal distribution. 
The probability distribution of the power in the fitted model 
at frequency $f_j$ is therefore:
\begin{equation}
p_{\hat{\mathcal{P}}_j}(y) = \frac{1}{S_j y \sqrt{2 \pi}} 
\exp \left\{ - \frac{(\ln[y] - M_j)^2}{2S_j^2}  \right\}
\end{equation}
where $M_j = \ln [\mathcal{P}_j ]$ and $S_j$ is the uncertainty
on the logarithm of the model (equation~\ref{eqn:model_err2}). 
The periodogram ordinate $I_j$ and the powe $\hat{\mathcal{P}}_j$
in the best fitting model are only strictly independent if the
frequency of interest, $f_j$, is excluded from the LS fit.
Otherwise the fitted model, and hence the model power at this
frequency, $\hat{\mathcal{P}}_j$,
is influenced by $I_j$. Although the effect is small ($\sim 1/n$) 
it does have a substantial impact on the tail of the PDF.
Thus, to obtain formally independent variables 
we must calculate $\hat{\mathcal{P}}_j$ (and its error)
using the LS method after ignoring $I_j$ from the fit.
In other words, we must ignore the candidate frequency when fitting
the continuum model and then compare the measured periodogram
ordinate at this frequency with the model derived from fitting all the 
other (independent) frequencies.

The PDF of the ratio $\hat{\gamma}_j$
can be obtained using the standard formula for the PDF 
of the ratio of two independent variables:
\begin{equation}
p_{\gamma_j}(z) = \int_{-\infty}^{+\infty} 
 |y| p_{2I_j}(zy) p_{\mathcal{P}_j}(y) dy
\end{equation}
The periodogram is always positive so we can integrate over
positive values only. 
\begin{equation}
p_{\gamma_j}(z) = 
\frac{1}{2 S_j \mathcal{P}_j  \sqrt{2 \pi}} 
\int_{0}^{+\infty} 
\exp \left\{ - \frac{(\ln[y] - M)^2}{2S_j^2} - \frac{zy}{2\mathcal{P}_j} \right\}
dy
\label{eqn:ratio_pdf}
\end{equation}
The dummy variable $w=y/P_j$ (and $dy=P_j dw$) can be used 
to simplify the above equation:
\begin{equation}
p_{\gamma_j}(z) = 
\frac{1}{S_j  \sqrt{8 \pi}} 
\int_{0}^{+\infty} 
\exp \left\{ - \frac{\ln[w]^2}{2S_j^2} - \frac{zw}{2} \right\}
dw
\label{eqn:ratio_pdf_simple}
\end{equation}
It is worth noting that this formula contains no dependence on the
actual value of $\mathcal{P}_j$ (and hence $M_j$). This
should not be surprising because we are dealing with the
distribution of the ratio of the data to the model, not
the absolute value of the data, and so the absolute value
of the model is not relevant. The important parameter
is $S_j$, describing the uncertainty on the model, and this
can be calculated from the theory of linear regression
(section~\ref{sect:error}). 
Unlike the ideal case discussed above (section~\ref{sect:ideal_stats}) 
the PDF is frequency-dependent,
this is reflected in the changes of $S_j$ with frequency.
As $S_j \rightarrow 0$ this formula reduces to the equivalent for
the ideal case discussed above.

\begin{figure}
\begin{center}
   \includegraphics[width=8 cm, angle=0]{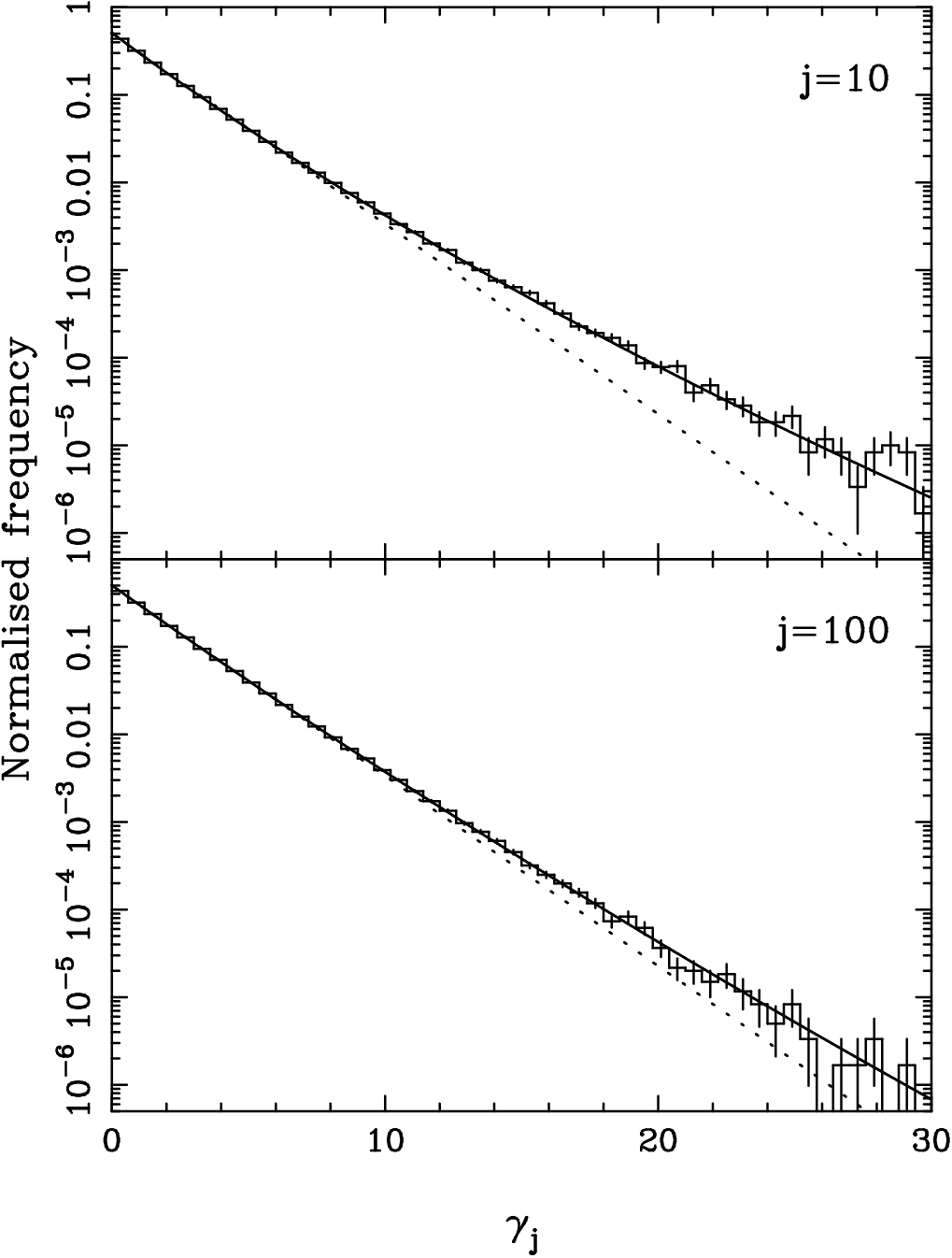}
\end{center}
\caption{
Monte Carlo demonstration of the PDF of the ratio $\hat{\gamma}_j$.
The histograms represent the distribution of the ratio
$\hat{\gamma}_j$ (measured at 
the $j=10$ and $j=100$ Fourier frequencies) from $10^6$ 
Monte Carlo realisations of $K=256$ time series (with an 
$\alpha=2, N=1$ spectrum).
For each of the $10^6$ simulated time series the periodogram was
fitted using the LS 
method, after ignoring frequency $j$, and the ratio $\hat{\gamma}_j =
2 I_j / \hat{\mathcal{P}}_j$ was measured.
The solid curve marks the predicted PDF using
equation~\ref{eqn:ratio_pdf_simple} and the dotted line marks
the PDF of a $\chi_2^2$ distribution (equation~\ref{eqn:chi2}).
The difference between the two model PDFs is due to the 
uncertainty on the model fit.
(Compare with Fig.~5 of Israel \& Stella \cite{is96}).
\label{fig:ratio_pdf}
}
\end{figure}

For a given frequency $f_j$ the integral in 
equation~\ref{eqn:ratio_pdf_simple} can be evaluated 
numerically to give the PDF for $\hat{\gamma}_j$.
Figure~\ref{fig:ratio_pdf} compares the prediction of
equation~\ref{eqn:ratio_pdf_simple} with a Monte Carlo distribution
at two different frequencies.
Also shown for comparison is the $\chi_2^2$ PDF which represents
the distribution in the absence of uncertainties on the model (i.e. $S_j \rightarrow 0$).
At small $\hat{\gamma}$ (i.e. low significance peaks)
the two distributions agree, whereas for large $\hat{\gamma}$ (i.e. high significance
peaks) there is a substantial difference between the PDFs including and
excluding the uncertainty on the model. This means that,
while the effect of including the uncertainty on the model is
negligible for low significance peaks, the significance of high
significance peaks may be substantially overestimated
if this additional uncertainty is not taken into account. 

The probability of obtaining a value of $\hat{\gamma}_j$ higher
than $\gamma_{\epsilon}$ can be computed by integrating this PDF:
\begin{equation}
\Pr \{ \hat{\gamma}_j > \gamma_{\epsilon} \} =
\int_{\gamma_{\epsilon}}^{\infty} p_{\gamma_j}(z) dz = \epsilon_1
\label{eqn:total_prob}
\end{equation}
This can be evaluated numerically to find $\gamma_{\epsilon}$ for a
given $\epsilon_1$.
Equivalently, we can find the value of $\gamma_{\epsilon}$ 
at the corresponding $1-\epsilon_1$ significance level:
\begin{equation}
\Pr \{ \hat{\gamma}_j < \gamma_{\epsilon} \} =
\int_{0}^{\gamma_{\epsilon}} p_{\gamma_j}(z) dz = 1 - \epsilon_1
\label{eqn:total_prob2}
\end{equation}
The calculation of $\gamma_{\epsilon}$ depends only on
$p_{\gamma_j}(z)$, from equation~\ref{eqn:ratio_pdf_simple}, 
which in turn depends only on  $S_j$, from equation~\ref{eqn:model_err2}, and
this is calculated using the the abscissae (frequencies $f_j$) with no
dependence on the ordinates (periodogram powers $I_j$).
The critical value $\gamma_{\epsilon}$ can be evaluated using only
the frequencies of the periodogram.

Finally, we need to correct for the number of frequencies examined.
The probability that a peak will be seen given that
$n^{\prime}$ frequencies were examined is $\epsilon_{n^{\prime}} = 1 -
(1-\epsilon_1)^{n^{\prime}}$. 
One can find the global $(1-\epsilon_{n^{\prime}})100$ per cent
confidence level by finding the value $\gamma_{\epsilon}$ that satisfies:
\begin{equation}
\int_{\gamma_{\epsilon}}^{\infty} p_{\gamma_j}(z) dz = 
1 - ( 1 - \epsilon_{n^{\prime}} )^{1/n^{\prime}} \approx
\epsilon_{n^{\prime}} / n^{\prime}
\label{eqn:conf_level}
\end{equation}
where $n^{\prime}$ is again the number of frequencies examined.

As an illustration of the effect of the model uncertainty, consider a
peak in the $j=10$ frequency bin of a $n=128$ 
periodogram. Neglecting the effect of model uncertainty  the nominal
$\epsilon_1=10^{-4}$ threshold is $\gamma_{\epsilon}=18.42$ (using
equation~\ref{eqn:1-error}). But, after including the uncertainty in the model,
the probability of this level being 
exceeded is really $3.6 \times 10^{-4}$ (using equation~\ref{eqn:total_prob}).  
For $n=128$ trials this corresponds to global
significances of $98.7$ per cent confidence (ignoring the model
uncertainty) and $95.4$ per cent confidence (including the model
uncertainty). 
The first of these might be called a significant
detection, but once the model uncertainty is taken into account
the detection is no longer very significant. The difference
is even more profound for higher significances.

% --------------------------------------------------------------------------
% --------------------------------------------------------------------------
% --------------------------------------------------------------------------
% --------------------------------------------------------------------------

\begin{figure}
\begin{center}
   \includegraphics[width=6.4 cm, angle=270]{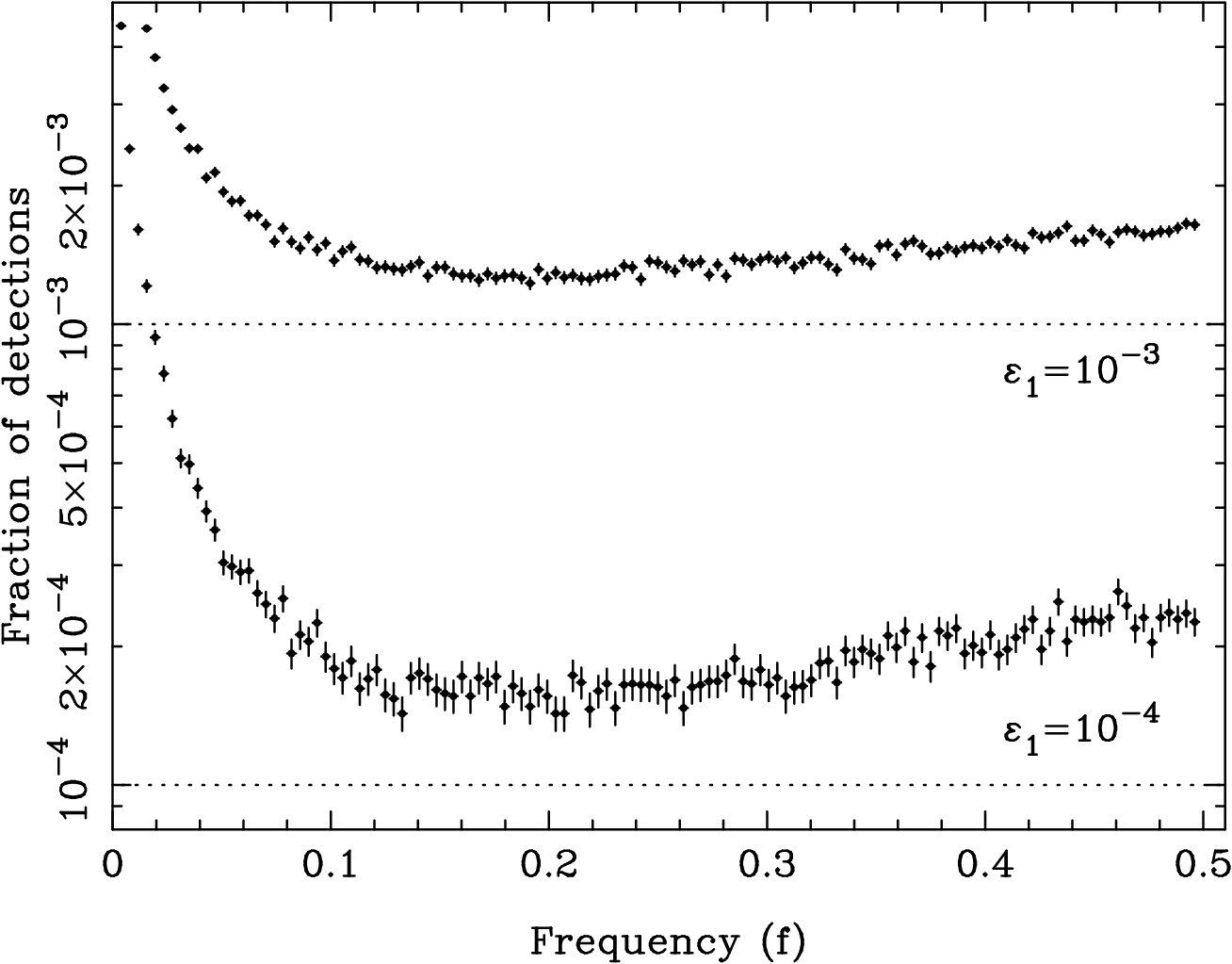}

\vspace{0.5 cm}
   \includegraphics[width=6.4 cm, angle=270]{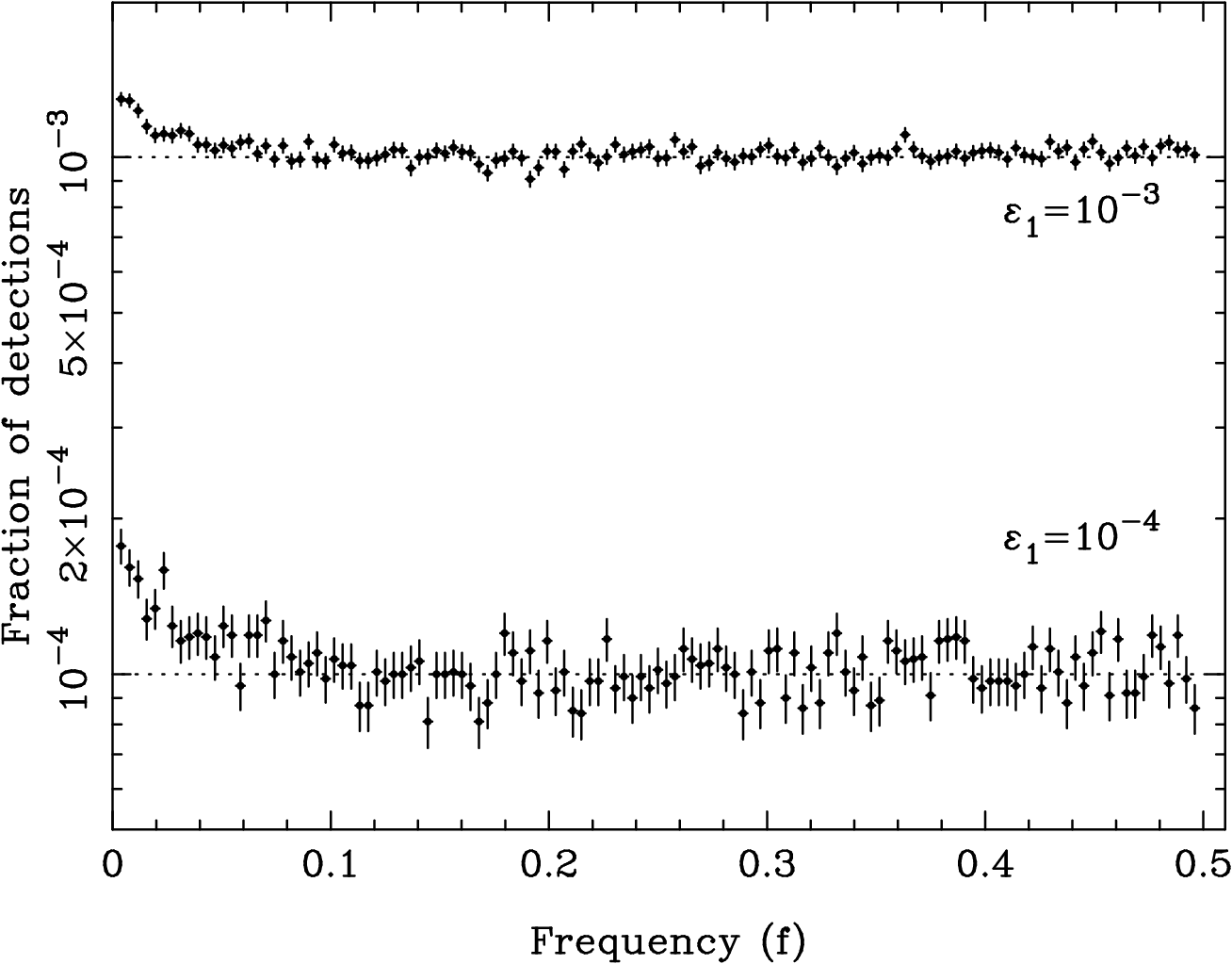}
\end{center}
\caption{
Monte Carlo study of the performance of tests for significant
periodogram peaks. For each panel $10^6$ random time series
of length $K=256$ were generated (with an $\alpha=2, N=1$ spectrum). 
For each of the $127$ frequencies tested ($=K/2-1$ since the Nyquist
frequency was ignored) the fraction of Monte Carlo simulations
with a peak exceeding the nominal $99.9$ and $99.99$ per cent 
confidence levels
was recorded. The upper panel used the confidence 
levels from equation~\ref{eqn:1-error}. The lower panel
used the confidence levels computed with equation~\ref{eqn:total_prob}
which accounts for the uncertainty in the model.
The error bars were computed using $\sqrt{p(1-p)/N}$ where $N$
is the number of simulations.
\label{fig:mc_error}
}
\end{figure}

\section{Verification of the method}
\label{sect:sims}

The procedures discussed above were tested using Monte Carlo
simulations. The simulations measured the type I error rate, or
the rate of `false positive' (spurious) detections of periodic signals.
For this experiment many artificial time series were generated based on a power law
spectrum (i.e. the null hypothesis). For each simulation the
$\gamma_{\epsilon}$ threshold, corresponding to a $1-\epsilon$
confidence level, was calculated and the number of
periodogram ordinates that exceeded this value were recorded.
The rate measured from the Monte Carlo simulations should be
the same as the false alarm probability $\epsilon$, often called the
`size of the test,' which is the expected rate of type I errors. 
If the observed rate of false detections
exceeds the nominal size of the test then one should expect an 
excess of spurious detections (detections may not be reliable). If the
observed rate falls below the nominal test size then the test is
conservative (it gives even fewer spurious detections than expected). 

The Monte Carlo rate was derived from $10^6$ random time series of
length $K=256$ (generated with a $\alpha=2, N=1$ spectrum).  For each
series the periodogram was computed and fitted using the LS method.
In the first run, the effects of the uncertainty on the model were
ignored and the $\gamma_{\epsilon}$ thresholds were computed for
$\epsilon_1 = 10^{-3}$ and $10^{-4}$ using equation~\ref{eqn:1-error}.  
These corresponds to $99.9$ and $99.99$ per cent confidence levels
in a single trial test. For each frequency the fraction of simulations that
show  peaks larger than the threshold was recorded. As shown in
Fig.~\ref{fig:mc_error} (upper panel) the observed rate of type I
errors in the simulated data was far in excess of the nominal size of the test.
Thus the actual rate of spurious detections was higher than
the nominal test size, and greatly so  at low frequencies where the
model is more uncertain.  The situation is worse at high significances
(small test sizes) because the tail of the PDF diverges from the
expectation (Fig.~\ref{fig:ratio_pdf}). This means that significances
calculated by equation~\ref{eqn:1-error} will be overestimated.

The situation is much better when the $\gamma_{\epsilon}$ threshold
was computed (again for $\epsilon_1 = 10^{-3}, 10^{-4}$) using 
equation~\ref{eqn:total_prob}. Again $10^6$ random time series were
generated with $K=256$ using the same spectrum. For each periodogram
the model power $\hat{\mathcal{P}}_j$, and its error parameters $S_j$, 
were computed by ignoring frequency $f_j$ and then fitting with the LS
method. The ratio $\hat{\gamma}_j$ was compared
to the critical threshold $\gamma_{\epsilon}$ (computed from
equation~\ref{eqn:total_prob}) at each frequency. The fraction of Monte Carlo 
simulations that showed `significant' peaks was very close
to the expected level (the nominal size) and independent of
frequency. The exceptions are the $\approx 5$ lowest frequencies.
Here the model is least constrained and the assumption
implicit in equation~\ref{eqn:model_err}, that the distribution
of $\log[\hat{\mathcal{P}}_j]$ is normal, becomes inaccurate.
For $j
{\mathrel{\hbox{\rlap{\hbox{\lower4pt\hbox{$\sim$}}}\hbox{$>$}}}} 5$
the confidence levels predicted by 
equation~\ref{eqn:total_prob} gave the correct rate of type I errors. 

\begin{figure}
\begin{center}
   \includegraphics[width=6.5 cm, angle=270]{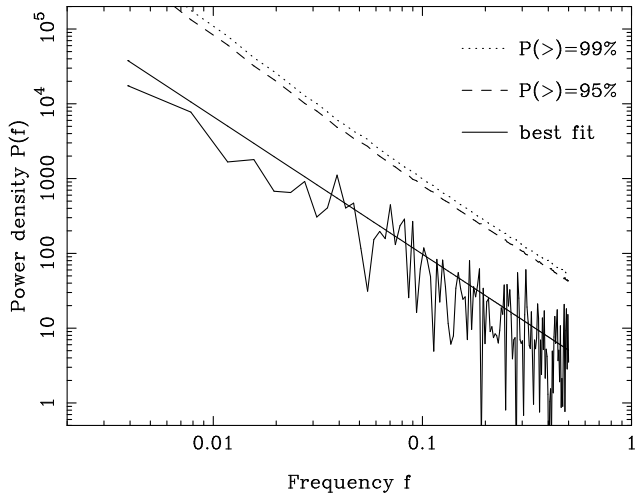}
\end{center}
\caption{
Same periodogram as in Fig.~\ref{fig:test_qpo_psd}.
Plotted are the de-biased LS estimate of the 
power law spectral model (solid curve) and the 
$95$ and $99$ per cent upper limits (dotted curves) on the expected
power (global significance levels for $n=127$ independent frequencies).
\label{fig:test_qpo_psd2}
}
\end{figure}

Figure~\ref{fig:test_qpo_psd2} shows a specific example, namely the
same data as in 
Fig.~\ref{fig:test_qpo_psd}, with the LS power law model.
Also shown are the (`global') $n^{\prime}$-trial confidence limits
computed as discussed in section~\ref{sect:model_err}.
Clearly none of the peaks in the periodogram exceeds the
$95$ per cent limit, as expected for red noise.

% --------------------------------------------------------------------------
% --------------------------------------------------------------------------
% --------------------------------------------------------------------------
% --------------------------------------------------------------------------

\section{Caveats and comparison with other methods}
\label{sect:caveats}

\subsection{Underlying assumptions}

In order for the test to give reliable significance limits  the
underlying noise spectrum must be a power law.   Clearly if the
broad-band noise spectrum does not resemble a power law the results of
the LS fitting will not be valid.  The general solution to this
problem is to replace the LS fitting  procedure with the exact
maximum likelihood (ML) procedure for fitting the $\chi_2^2$ distributed
periodogram. The appendix describes this method. 

The test was intended to be used for assessing the significance of
peaks in the periodograms of X-ray observations of Seyfert galaxies,
which tend to be rather short $K \sim 10^3$ and also show significant
variance due to measurement errors (Poisson noise). These measurement
uncertainties produce a flat component that is added to the source
power spectrum in the  periodogram.  The effect will cause the
observed spectrum to flatten at high frequencies as the power in the
red noise spectrum of the source becomes  comparable to the power in
the flat Poisson noise spectrum. Using the normalisation given in
equation~\ref{eqn:ft2} the expected Poisson noise level is $P_N =
2(\langle x \rangle + B)/ \langle x \rangle ^2$ where $\langle x
\rangle $ is the mean count rate and $B$
is the mean background rate\footnote{
For the case of Gaussian errors $\sigma_k$ on each flux measurement
the expected Gaussian noise level is $P_N = 2 \Delta T \langle
\sigma^2 \rangle
/ \langle x \rangle^2$. This assumes the time series comprises contiguous
bins of length $\Delta T$.}. 
It is also now known that at low frequencies the power spectra
 of Seyferts break from a single power law (e.g. Uttley et al. 2002;
 Markowitz et al. 2003). These deviations from a single power law
 should be accounted for in modelling the spectrum.
 The simplest solution is to divide the periodogram into 
 frequency ranges within which the power spectrum is approximately
 a single power law. The period detection test can then be used
 as described above.
 The crucial point is that as long as the periodogram can be fitted
 reasonably well with a power law over a frequency
 range of interest (as judged using the KS test) the test
 will be valid.
Alternatively one may fit a model of a power law plus
constant (to account for the flattening) using the ML
method discussed in the Appendix.

Furthermore, the test, which is based on the discrete Fourier transform, is
most sensitive to sinusoidal periodicities. Non-sinusoidal
variations will have their power spread over several frequencies
which will lessen the detection significance in any one given
frequency. Other methods such as epoch folding (Leahy et
al. \cite{lea}), whereby one
bins the time series into phase bins at a
test period, can be more sensitive to such variations.
At the correct period the periodic variations will
sum while any background noise will cancel out, thus revealing
the profile of the periodic pulsations. However, the background
noise will only cancel out if it is temporally independent, i.e.
white noise. Again, the presence of any underlying red noise
variations may produce unreliable results (Benlloch et
al. \cite{ben}) if not correctly accounted. 

% --------------------------------------------------------------------------

\subsection{Comparison with other methods}

\subsubsection{Lomb-Scargle periodogram}

The method described above is based around the standard
(Fourier) periodogram and therefore requires uniformly sampled time
series. This ensures the asymptotic independence of the periodogram
ordinates. If the time series is non-uniformly sampled one may use
other periodogram estimators such as the Lomb-Scargle periodogram
(Lomb \cite{lomb}; Scargle \cite{sca82}; Press \& Rybicki
\cite{pr89}).   However, the behaviour of these will not be identical
to that discussed above. The above procedure should not be used on
non-uniformly sampled time series (nor should the method discussed in
section 13.8 of Press et al. \cite{press96} be used in the presence of
non-white noise).  Zhou \& Sornette (\cite{zs}) discuss the results of
Monte Carlo tests on the distribution of peaks in the Lomb-Scargle
periodogram for various types of processes.

\subsubsection{Oversampled periodogram}
\label{sect:oversample}

Oversampling the periodogram, i.e. calculating periodogram
ordinates at frequencies between the normal Fourier frequencies,
is sometimes done in order to increase the sensitivity to weak
signals that lie at frequencies between two Fourier frequencies.
A periodic variation with a frequency nearly equidistant
between two adjacent Fourier frequencies, e.g. between $f_j$ and $f_{j+1}$, 
will have its power spread (almost entirely) between these two 
frequencies, thus reducing the significance in any one frequency bin. 
The reduction in power per bin can be as much as $\approx 4/\pi^2 \approx 0.41$. 
In these situations oversampling the periodogram by including additional
frequencies between $f_j$ and $f_{j+1}$ can increase sensitivity
to the periodicity. However, it must also be noted that by
oversampling the periodogram one is testing more than $K/2$ frequencies,
allowing many more opportunities to find spurious peaks. The number of
trials increases above the usual $n=K/2$ case if the periodogram is
oversampled, although the effective number of independent trials does
not scale linearly with the 
oversampling factor because there is a fixed number ($n$) of
strictly independent frequencies (the Fourier frequencies).

\begin{figure}
\begin{center}
   \includegraphics[width=6.5 cm, angle=270]{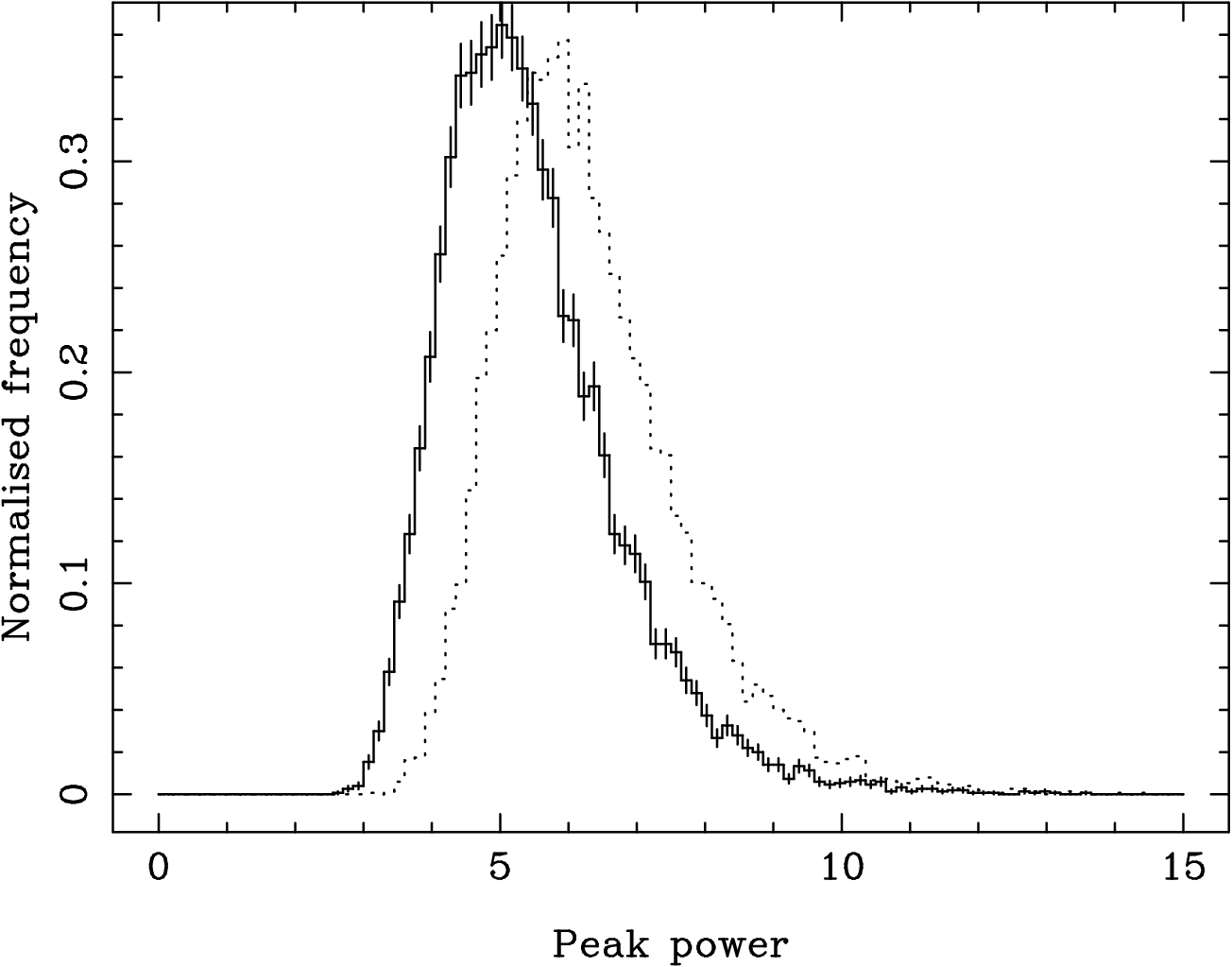}

\vspace{0.5 cm}
   \includegraphics[width=6.5 cm, angle=270]{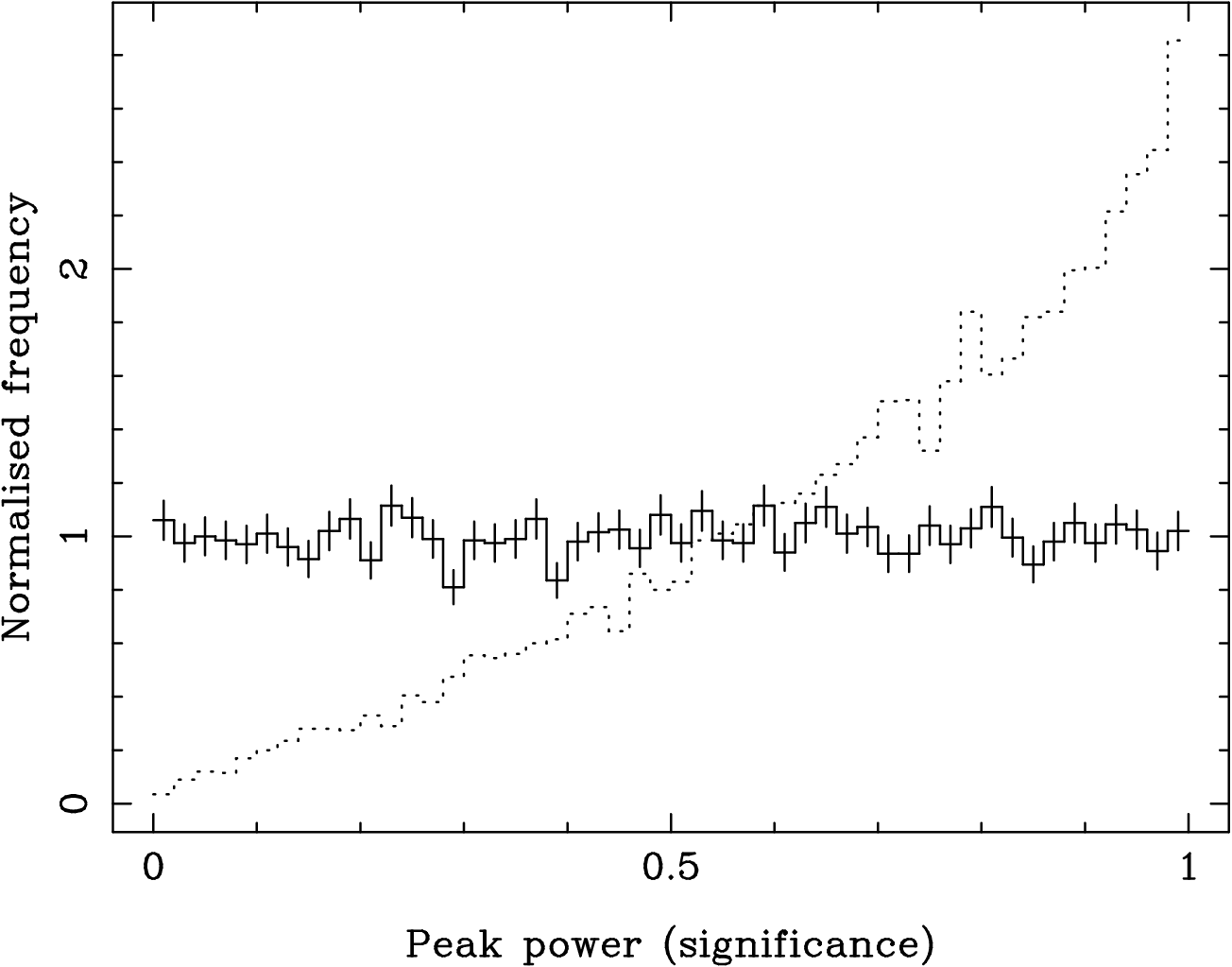}
\end{center}
\caption{
Comparison between periodograms sampled at the $K/2$ Fourier
frequencies and oversampled by a factor $8$. The upper panel shows
the distribution of peaks in the periodogram from $10^4$
random white noise time
series of length $K=256$ (using an $\alpha=0, N=1$ spectrum).
For each simulation the value of the peak of the periodogram was recorded for the
standard (solid histogram) and oversampled (dotted histogram) case.
The oversampled periodograms clearly show slightly larger peak values. 
The lower panel shows the same data except the peak powers have
been converted to `global' significances using equation~\ref{eqn:global}.
\label{fig:oversample}
}
\end{figure}

Oversampling the periodogram and assuming $\approx n$ trials will
tend to
overestimate the significance of peaks in the periodogram.
This was demonstrated by Monte Carlo simulations (see
Fig.~\ref{fig:oversample}). For this demonstration 
$10^4$ white noise  time series (spectrum: $\alpha=0, N=1$)
were generated with length $K=256$ and for each one the periodogram
was calculated using both the standard Fourier frequencies
and also oversampling the frequency resolution by a factor $8$. 
The peak power in the periodogram of  each of the simulated time series
was recorded. The distribution of peak powers is shown in
Fig.~\ref{fig:oversample} for both the standard and the oversampled
periodograms. There is an obvious tendency for the oversampled
periodogram to show slightly higher peak values, as might be expected
based on the above arguments. These peak powers were translated to 
`global' significances (assuming $n$ independent trials) and the
distribution of significances is shown in the lower panel of
Fig.~\ref{fig:oversample}. The distribution is flat 
for the Fourier sampled periodogram: the significance of the peaks
is exactly as expected. The distribution derived from the oversampled 
periodogram clearly shows a substantial excess of high significance
peaks. This means that the oversampled periodogram is likely to
produce many more spurious peaks than expected if one assumes only
$n$ independent trials were made. 
A Monte Carlo estimate of the {\it global} significance would be
required to calibrate the significance of peaks in oversampled
periodogram (and find a more realistic effective number of trials).

An alternative to oversampling the periodogram is to perform
a sliding two-bin search for periodogram peaks using the
standard Fourier frequencies only, as discussed by
van der Klis (\cite{van89}; section 6.4). This increases the sensitivity to
periods whose frequencies fall between the Fourier frequencies.
However, one will still need to perform Monte Carlo simulations
to assess the global significance (by measuring the rate of type I
errors in the simulations). 

\subsubsection{Monte Carlo testing}

An alternative test for periodic variations in red noise is to
estimate the likelihood of observing a given peak using Monte Carlo
simulations of red noise processes (Benlloch et al. \cite{ben};
Halpern, Leighly \& Marshall \cite{hlm}). However, the method of
Benlloch et al. (\cite{ben}) does not account for uncertainties in the
best-fitting model, which can seriously effect the apparent
significances of strong peaks (section~\ref{sect:model_err}). Monte
Carlo simulations are only  as good as the model they assume! One
solution would be to map the (multi-dimensional) distribution of the
model parameters and for each simulation draw the model 
 parameters at random from 
this distribution. This would  thereby account for the uncertainty in
the model parameters.   (Protassov et al. \cite{prot} use a similar
approach to calculate posterior predictive $p$-values.)

% --------------------------------------------------------------------------
% --------------------------------------------------------------------------
% --------------------------------------------------------------------------
% --------------------------------------------------------------------------

\section{Discussion}
\label{sect:disco}

A simple procedure is presented for
assessing the significance of 
peaks in a periodogram when the underlying
continuum noise has a power law spectrum. 

\subsection{Recipe}

The following is one possible recipe for periodogram analysis.

\begin{itemize}
\item
Calculate the periodogram of the data $I(f_j)$
(section~\ref{sect:per}) and convert to log-space, i.e. $\log [ f_j ]$
and $\log [ I(f_j) ]$. 

\item
Ignore the Nyquist frequency
and frequencies above which Poisson noise is significant,
leaving $n^{\prime}$ frequencies.

\item 
Estimate the power law parameters by fitting
a linear function to the log-periodogram using
the LS method (section~\ref{sect:fitting}). 

\item
Test the goodness of the fit by comparing the distribution
of data/model residuals with the $\chi_2^2$ expectation
using a KS test (section~\ref{sect:fitting}). 

\item
Calculate the `global' ($n^{\prime}$-trial) threshold ratio for
a $(1-\epsilon)$ significance detection using
$\gamma_{\epsilon} = - 2 \ln [ 1 - ( 1 - \epsilon )^{1/n^{\prime}} ]$
where $n^{\prime}$ 
is number of periodogram points used in the fit
and $\epsilon$ is the desired `false alarm probability'
(section~\ref{sect:ideal_stats}).

\item
Multiply the best-fitting continuum model by $\gamma_{\epsilon}/2$ (or
add its logarithm in log-space).
\end{itemize}

This will provide an estimate of the power spectral slope
and normalisation and a first indication of the presence of 
significant periodicities. However, as discussed in
section~\ref{sect:model_err} the uncertainty inherent in the model
fitting means that the significances of strong peaks may be
substantially overestimated. This procedure should be quite accurate
for low significance peaks however, meaning that the above procedure is
valid for rejecting low significance peaks ($\epsilon_1 > 0.01$). 
The inaccuracy of the method at high powers
means that the procedure should only be used to reject low
significance peaks, not detect high significance peaks.
In order to obtain a more rigorous estimate of the significance of 
stronger peaks one needs to account for the additional uncertainty in
the model. 

For a given frequency of interest, $f_j$, one must: 

\begin{itemize}

\item
Ignore $f_j$ from the fit (in order that the data and model
are independent at $f_j$).

\item
Re-fit the periodogram using the LS method (section~\ref{sect:fitting}).

\item
Calculate the power in the model at this frequency, $\hat{\mathcal{P}}_j$.

\item
Measure the ratio $\hat{\gamma}_j = 2 I_j / \hat{\mathcal{P}}_j$.

\item
Compute the uncertainties on the model
parameters (section~\ref{sect:error}) and hence the
uncertainty on the model continuum $S_j$ (section~\ref{sect:error2}). 

\item
Calculate the probability of observing a value of $\hat{\gamma}_j$
this high by numerically integrating equation~\ref{eqn:total_prob}.
\end{itemize}

The above procedure is essentially the application of two standard tools
for time series analysis.  The first is estimating the power law
spectral properties by LS fitting of the
log-periodogram. This has been discussed by  several authors
(e.g. Geweke \& Porter-Hudak \cite{gph}; Fougere \cite{fou}; Pilgram
\& Kaplan \cite{pil}).  The second tool is applying the known
$\chi_2^2$ properties of the periodogram to estimate confidence
levels, as is standard for white noise spectra (Priestley
\cite{pri81}; van der Klis \cite{van89}). 
The additional calculation to include the uncertainty in
the model follows standard statistical procedures.
The overall approach is similar to that discussed by
Israel \& Stella (\cite{is96}) but is tailored to
power law spectra.

 In order for the method to produce reliable results the underlying
 power spectrum has to be a power law. However, the test
 will work well even for data that show deviations from 
 a power law (such as intrinsic low frequency flattening 
 or high frequency flattening due to Poisson noise) provided
 the periodogram is divided into frequency intervals within
 which a single power law provides a good description of the
 data (determined with a KS test). In this case the power spectrum
 over the restricted frequency ranges is indistinguishable from 
 a power law and, when applied to these limited frequency ranges,
 the method will function as expected.

% --------------------------------------------------------------------------

\subsection{Application to real data}

\begin{figure}
\begin{center}
   \includegraphics[width=6.5 cm, angle=270]{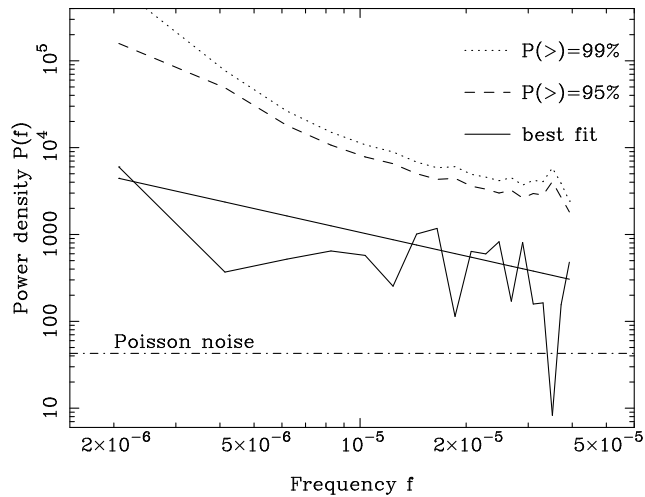}\\
\end{center}
\caption{
Periodogram of IRAS 18325--5926 from {\it ASCA}.
Plotted are the de-biased LS estimate of the 
power law spectral model (solid curve) and the 
$95$ and $99$ per cent upper limits (dotted curves) on the expected
power (global significance levels for $n=19$ independent
frequencies). Also shown is the expected level of the Poisson noise
power. 
\label{fig:test1}
}
\end{figure}

As a demonstration of the method we present a re-analysis of the X-ray
observations of two Seyfert galaxies.  The first is the long {\it
ASCA} observation of IRAS 18325--5926  (Iwasawa et al. \cite{iwa}) and
the second is the {\it XMM-Newton} GTO observation of Mrk 766 (Boller
et al. \cite{bol}). For IRAS 18325--5926  a background-subtracted
$0.5-10$~keV SIS0 light curve  was extracted in $100$-s bins, using a
$4$~arcmin radius source region, and rebinned onto an evenly spaced
grid at the  spacecraft orbital period ($5760$-s). For Mrk 766  a
background-subtracted, exposure-corrected $0.2-2$~keV light curve was
extracted  from the EPIC pn using a $38$~arcsec radius source
region. The rebinned IRAS 18325--5926 light curve contained $84$
regularly spaced bins and the Mrk 766 light curve contained $330$
contiguous $100$-s bins.

\begin{figure}
\begin{center}
   \includegraphics[width=6.5 cm, angle=270]{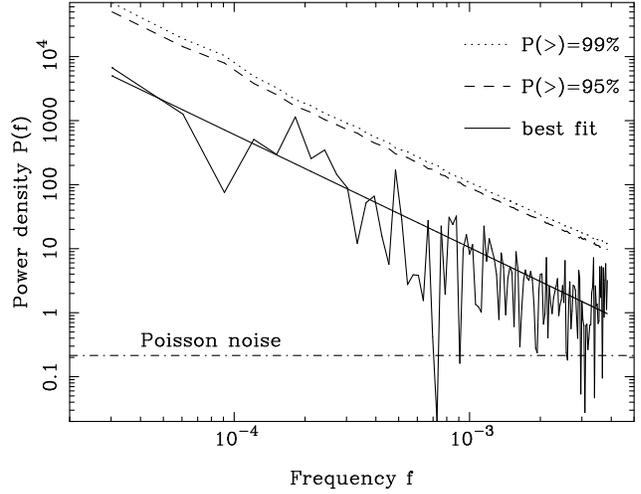}
\end{center}
\caption{
Periodogram of Mrk 766 from {\it XMM-Newton}.
\label{fig:test2}
}
\end{figure}

The periodogram of each light curve was calculated and the expected
Poisson noise level determined.  Only the
lowest frequency periodogram points were examined, to minimise the
effect of the Poisson noise level. For IRAS 18325--5926 only the $19$
lowest frequency points were used, while for Mrk 766 only $97$ points
were used. These were fitted with a linear function in log-space,
giving slopes of $\alpha \approx 0.9$ and $\alpha \approx 1.8$ for
IRAS 18325--5926 and Mrk 766, respectively.  The $95$ and $99$ per
cent confidence limits on the periodogram  were computed, accounting
for the number of frequencies examined in each case (see
figures~\ref{fig:test1} and \ref{fig:test2}).  In neither object did a
single periodogram point exceed the $95$ per cent limit, meaning that
there is no strong evidence to suggest a periodic component to their
variability, contrary to the original claims of Iwasawa et
al. (\cite{iwa}; $f_P \approx 1.7 \times 10^{-5}$~Hz) and Boller et
al. (\cite{bol}; $f_P \approx 2.4 \times 10^{-4}$~Hz). Benlloch et 
al. (\cite{ben}) drew similar conclusions based on Monte Carlo
simulations. 

% --------------------------------------------------------------------------

\subsection{Conclusions}

The simplicity of the proposed test means that it can be used as a
quick, first check against spurious periodogram peaks, without the
need for extensive Monte Carlo simulations.  
Examples where this test might be useful include
not only X-ray observations of Seyferts but  also testing for periodic
components in monitoring of blazars (e.g. Kranich et al. \cite{kra};
Hayashida et al. \cite{hay}), the Galactic centre (Genzel et
al. \cite{genz}) and other astrophysical sources that show red noise
variations.

The basic idea behind the
method is to de-redden the periodogram by dividing out the
best-fitting power law and then using the known distribution of
the periodogram ordinates to estimate the likelihood of observing a
given peak if the null hypothesis (power law spectrum with no
periodicity) is true.  However, this should not be treated as a `black
box' solution to the problem of detecting periodicities in red noise
data. There is no substitute for a thorough understanding of the
nuances of  power spectral statistics, as illustrated by the PDFs
discussed in section~\ref{sect:model_err}. The tail of the PDF is
very sensitive to the details of the method, and thus one needs to
treat any periodogram analysis with great care in order to avoid
overestimating the significance of a peak.

To date the most significant candidate periodicities found
in the X-ray variations active galaxies
are those in long {\it EUVE} observations of three nearby
Seyferts by Halpern et al. (\cite{hlm}). However, their
published global significances will be overestimates for two 
reasons: ($i$) no account was made of the uncertainty
on the model and ($ii$) the periodograms were oversampled.
These two separate issues both act to artificially boost the
apparent significance of spurious signals, as discussed in
sections~\ref{sect:model_err} and \ref{sect:oversample},
respectively. 
Similarly the apparently significant candidate periodicity in NGC 5548
found by Papadakis \& Lawrence (\cite{pap2}) was shown 
to be substantially less significant when the uncertainty in the
modelling was included (Tagliaferri et al. \cite{tag}). 
This leaves us in the interesting situation
of there being not one surviving, robustly determined and
significant ($>99$ per cent) periodicity in the X-ray variability of
an active galaxy. 

Finally we close with a plea. There exist many more AGN light curves
and periodograms than have been published.  There is a natural
publication bias: those that show the most `significant' features get
published. This means the true number
of trials is much larger than for any one given experiment. In such
situations one should treat 
with caution the significance of the subset of results that are
published (e.g. Scargle \cite{sca00}). 
That is, until a result is published that is so significant
that it cannot be accounted for  by publication bias.  The detection
of periodic or quasi-periodic variations in galactic nuclei would be a
major discovery and of great importance to the field.  The importance
of the result should be argument enough for very high standards of
discovery.  We would therefore advocate serious further investigation of
only those candidate periodicities with high significances 
(such as a $>99.9$ per cent confidence, or a ``$3 \sigma$ minimum,''
criterion) after accounting for all likely sources of error.

\begin{acknowledgements}
I acknowledge financial support from the PPARC, UK.
I also thank Phil Uttley for many useful discussions 
and Iossif Papadakis for a very thorough referee's report
that encouraged me to expand several sections of the paper.
\end{acknowledgements}

% --------------------------------------------------------------------------
% --------------------------------------------------------------------------
% --------------------------------------------------------------------------
% --------------------------------------------------------------------------

\appendix

\section{Maximum likelihood (ML) fitting}
\label{sect:app}

In this section we briefly elucidate the exact method of
maximum likelihood (ML) fitting periodograms. See Anderson, Duvall \&
Jefferies (\cite{and}) and Stella et al. (\cite{ste}) for more
details. The probability density function (PDF) of the periodogram ordinates
at each frequency are given by:
\begin{equation}
p(I_j) = \frac{1}{\mathcal{P}_j}
           {\rm e}^{-I_j /  \mathcal{P}_j }
\end{equation}
where $\mathcal{P}_j$ is the true underlying spectrum
(the expectation value of the periodogram) at frequency $f_j$. 
This is valid for frequencies $j=1,2,\ldots,n-1$ since the periodogram
ordinate at the $j=n$ (Nyquist) frequency is distributed differently.

Assuming a model $\hat{\mathcal{P}}_j (\hat{\theta}_k)$, determined by
parameters $\hat{\theta}_k = \{ \hat{\theta}_1, \hat{\theta}_2, \ldots, \hat{\theta}_M
\}$, we can write the joint
probability density of observing the $n-1$ periodogram ordinates:
\begin{equation}
\mathcal{L} = \prod_{j=1}^{n-1} p(I_j) =  \prod_{j=1}^{n-1}
\frac{1}{\hat{\mathcal{P}}_j} {\rm e}^{-I_j / \hat{\mathcal{P}}_j } 
\end{equation}
If the data $I_j$ have already been observed this
represents the likelihood function. 
Maximising the likelihood $\mathcal{L}$ is 
equivalent to minimising $S \equiv - 2 \ln [ \mathcal{L} ]$ and we can rewrite this as:
\begin{equation}
S = 2 \sum_{j=1}^{n-1} \left\{ \ln[\hat{\mathcal{P}}_j] + \frac{I_j}{\hat{\mathcal{P}}_j} \right\}
\end{equation}
Finding the model parameters $\hat{\theta}_k$ that minimise $S$
will yield the maximum likelihood parameter values.

One can then use standard tools of maximum likelihood analysis, 
such as the likelihood ratio test (LRT) to test for additional
free parameters in the model:
\begin{equation}
R = - 2 \ln [\mathcal{L}_1 / \mathcal{L}_2] = S_1 - S_2 
\end{equation}
where $\mathcal{L}_1$ represents the likelihood for the simpler
(more parsimonious) model and $\mathcal{L}_2$ represents the 
likelihood for the model with the additional free
parameters. The second, more complex model, always contains 
within it the simpler model as a subset (the models are nested)
and therefore the likelihood for the more complex model is always
equal to or greater than that for the simpler model.
With certain restrictions (see Protassov et
al. \cite{prot}) $R$ is distributed as a $\chi_{\nu}^{2}$ variable
where $\nu$ is the number of additional free parameters. 
This test could be used, for example, to compare nested models such as
broken and unbroken power laws. 

Alternatively one can compare different models using the Akaike
Information Criterion ($AIC$; Akaike \cite{aka}):
\begin{equation}
AIC_i = - 2 \ln [ \mathcal{L}_i ] + 2 k_i = S_i + 2 k_i
\end{equation}
where $\mathcal{L}_i$ is the likelihood and $k_i$ is the number of
free parameters for model $i$. The second term in the sum is
a penalty for including more free parameters. The model
that minimises the $AIC$ is considered to be the best
(the models need not be nested).

One can also use $\Delta S = -2\Delta \ln[\mathcal{L}]$ to place
confidence limits on the model parameters in a fashion exactly
analogous to mapping confidence contours using $\Delta \chi^2$
(section 15.6 of Press et al. \cite{press96}).  Under fairly general conditions
(see Cash \cite{cash}), e.g. the $\ln[\mathcal{L}]$-surface is
approximately shaped like a multi-dimensional paraboloid, $\Delta S$
distributed as $\chi_{\nu}^{2}$ where $\nu$ is the  number of
parameters of interest (e.g. $\nu=1$ for the one-dimensional
confidence region on an individual parameter). One can use 
standard tables of $\chi_{\nu}^2$ values to place
confidence limits (e.g. $\Delta S = 2.71$ corresponds to $90$ per cent
confidence limits on one parameter). 

For the purposes of period searching one may fit a suitable
$M$-parameter continuum model (representing the null hypothesis, i.e.,
no periodic signal) using the ML method and define the
$M$-dimensional distribution of  its parameters (using
$\Delta S$).  One can then use this $M$-dimensional
distribution of the model  parameters to randomly draw models for
Monte Carlo simulation.  This procedure will thereby account for the
likely distribution of model parameters (which gives rise to
uncertainties in the estimated continuum level).

% --------------------------------------------------------------------------
% --------------------------------------------------------------------------
% --------------------------------------------------------------------------
% --------------------------------------------------------------------------

\end{document}